\documentclass[twocolumn]{aastex631}

\usepackage[english]{babel} 
\usepackage[utf8]{inputenc} 
\usepackage[T1]{fontenc}    
\DeclareUnicodeCharacter{0301}{/}  
\DeclareUnicodeCharacter{2212}{-}  
\usepackage{ae,aecompl}
\usepackage{pgf,pgfarrows,pgfnodes,pgfautomata,pgfheaps}
\usepackage{graphicx}       
\usepackage{natbib}         
\usepackage{url}            
\usepackage{grffile}        
\usepackage{orcidlink}
\usepackage{mathtools}      
\usepackage{multirow}       
\usepackage{xspace}         

\usepackage{txfonts}
\usepackage{amsmath,amssymb,amsxtra,amsfonts}   

\usepackage{color}         
\definecolor{gold}{rgb}{1,0.80,0}
\definecolor{orange}{rgb}{1,0.5,0}
\definecolor{midgray}{gray}{0.3}
\definecolor{lblue}{rgb}{0,0.2,0.5}
\definecolor{dgreen}{rgb}{0.1,0.6,0.3}
\definecolor{purple}{rgb}{0.5019607843137255,0.0,0.5019607843137255}

\renewcommand{\arcdeg}{\mbox{$^{\circ}$}\xspace}             

\newcommand{\be}{\begin{equation}}
\newcommand{\ee}{\end{equation}}

\newcommand{\ba}{\begin{align}}
\newcommand{\ea}{\end{align}}

\newcommand{\defeq}{\vcentcolon=}

\newcommand{\Mstar}{\ensuremath{M_\ast}\xspace}

\newcommand{\K}{\ensuremath{\rm K}\xspace}

\newcommand{\Hunit}{\ensuremath{\rm km~s^{-1}~Mpc^{-1}}\xspace}

\newcommand{\Hb}{\textrm{H}\ensuremath{\beta}\xspace}
\newcommand{\Hg}{\textrm{H}\ensuremath{\gamma}\xspace}
\newcommand{\Hd}{\textrm{H}\ensuremath{\delta}\xspace}

\newcommand{\OII}{[\textrm{O}\textsc{ii}]\xspace}
\newcommand{\OIII}{[\textrm{O}\textsc{iii}]\xspace}

\newcommand{\NII}{[\textrm{N}\textsc{ii}]\xspace}

\newcommand{\NeIII}{[\textrm{Ne}\textsc{iii}]\xspace}

\def\U{\ensuremath{U_{360}}\xspace}     
\def\B{\ensuremath{B_{435}}\xspace}
\def\V{\ensuremath{V_{606}}\xspace}
\def\I{\ensuremath{I_{814}}\xspace}
\def\Y{\ensuremath{Y_{105}}\xspace}
\def\J{\ensuremath{J_{125}}\xspace}

\def\H{\ensuremath{H_{160}}\xspace}

\def\p{{\rm prior}}

\newcommand{\Om} {\ensuremath{\Omega_{\rm{m}}}\xspace}

\newcommand{\Ol} {\ensuremath{\Omega_{\Lambda}}\xspace}

\newcommand{\n}{\ensuremath{{\nu}\rm}\xspace}

\usepackage{orcidlink}
\usepackage{booktabs} 
\usepackage{soul}
\usepackage{CJK}

\graphicspath{{./}{figures/}}
\shorttitle{Gas-phase Metallicity Gradients in Protoclusters at $z\sim2$}
\shortauthors{Yi-Ming Yang et al.}

\begin{document}
\begin{CJK*}{UTF8}{gbsn}

\title{MAMMOTH-Grism: Gas-phase Metallicity Gradients of Star-forming Galaxies in Protocluster Environments at Cosmic Noon}

\correspondingauthor{Xin Wang}
\email{xwang@ucas.ac.cn}

\author[0000-0002-0663-814X]{Yi-Ming Yang}
\affiliation{National Astronomical Observatories, Chinese Academy of Sciences, Beijing 100101, China}
\affiliation{School of Astronomy and Space Science, University of Chinese Academy of Sciences (UCAS), Beijing 100049, China}

\author[0000-0002-9373-3865]{Xin Wang}
\affiliation{School of Astronomy and Space Science, University of Chinese Academy of Sciences (UCAS), Beijing 100049, China}
\affiliation{National Astronomical Observatories, Chinese Academy of Sciences, Beijing 100101, China}
\affiliation{Institute for Frontiers in Astronomy and Astrophysics, Beijing Normal University,  Beijing 102206, China}

\author[0000-0002-9390-9672]{Chao-Wei Tsai}
\affiliation{National Astronomical Observatories, Chinese Academy of Sciences, Beijing 100101, China}
\affiliation{Institute for Frontiers in Astronomy and Astrophysics, Beijing Normal University,  Beijing 102206, China}
\affiliation{School of Astronomy and Space Science, University of Chinese Academy of Sciences (UCAS), Beijing 100049, China}

\author[0000-0001-5951-459X]{Zihao Li}
\affiliation{Cosmic Dawn Center (DAWN), Denmark}
\affiliation{Niels Bohr Institute, University of Copenhagen, Jagtvej 128, DK2200 Copenhagen N, Denmark}
\affiliation{Department of Astronomy, Tsinghua University, Beijing 100084, China}

\author[0000-0001-8467-6478]{Zheng Cai}
\affiliation{Department of Astronomy, Tsinghua University, Beijing 100084, China}

\author[0000-0002-8630-6435]{Anahita Alavi}
\affiliation{Infrared Processing and Analysis Center, Caltech, 1200 E. California Blvd., Pasadena, CA 91125, USA}

\author[0000-0002-1620-0897]{Fuyan Bian}
\affiliation{European Southern Observatory, Alonso de Cordova 3107, Casilla 19001, Vitacura, Santiago 19, Chile}

\author[0000-0001-6482-3020]{James Colbert}
\affiliation{Infrared Processing and Analysis Center, Caltech, 1200 E. California Blvd., Pasadena, CA 91125, USA}

\author[0000-0003-3310-0131]{Xiaohui Fan}
\affiliation{Steward Observatory, University of Arizona, 933 North Cherry Ave., Tucson, AZ 85721, USA}

\author[0000-0002-6586-4446]{Alaina L. Henry}
\affiliation{Space Telescope Science Institute, 3700 San Martin Dr., Baltimore, MD, 21218, USA}

\author[0000-0001-6919-1237]{Matthew A. Malkan}
\affiliation{Department of Physics and Astronomy, University of California, Los Angeles, 430 Portola Plaza, Los Angeles, CA 90095, USA}

\author[0000-0002-3264-819X]{Dong Dong Shi}
\affiliation{Center for Fundamental Physics, School of Mechanics $\&$ Optoelectronic Physics, Anhui University of Science and Technology, Huainan 232001, China}

\author[0000-0002-7064-5424]{Harry I. Teplitz}
\affiliation{Infrared Processing and Analysis Center, Caltech, 1200 E. California Blvd., Pasadena, CA 91125, USA}

\author[0000-0003-3728-9912]{Xian~Zhong Zheng}
\affiliation{Tsung-Dao Lee Institute and State Key Laboratory of Dark Matter Physics, Shanghai Jiao Tong University, Shanghai 201210, China}

\begin{abstract}
Environment plays a crucial role in shaping galaxy formation, yet the impact of overdensities on the internal chemical structure of galaxies at cosmic noon is still under debate.
Here, we present spatially resolved gas-phase metallicity gradients for 42 star-forming galaxies in three massive protoclusters at $z \sim 2.3$, derived from \textit{Hubble Space Telescope (HST}) slitless grism spectroscopy from the MAMMOTH-Grism survey.
We find that the majority (29 of 42, $\sim$69\%) of these protocluster members exhibit positive (inverted) metallicity gradients, a fraction significantly higher than observed in field galaxies of similar mass and redshift.
By examining correlations with global properties, we show that these positive gradients are strongly associated with galaxies that are metal-deficient relative to the field mass-metallicity relation, particularly among the massive population ($\log(M_*/M_\odot) > 9.95$).
These trends suggest that galaxies in dense protocluster environments experience substantial, enhanced inflows of pristine gas toward their central regions, which dilute the central metallicity and produce the observed inverted gradients.
Our results provide observational evidence that environmental effects actively regulate gas accretion and chemical redistribution during the peak epoch of cosmic star formation.
\end{abstract}

\keywords{High-redshift galaxies --- Galaxy environments --- Chemical abundances	 --- Metallicity --- Galaxy evolution	} 


\section{Introduction} \label{sec:intro}
The gas-phase metallicity (hereafter metallicity) is one of the crucial observational probes in tracing the baryonic cycle in galaxies,
which can provide key constraints on galaxy formation and evolution. It can be affected by star formation history, gas inflow, outflow, recycling and some other physical processes.
Beyond the global abundance, the spatial distribution of metallicity within galaxies,
i.e., the metallicity gradient, can further reveal the internal physical processes that form and shape galaxies.    
In an ideal inside-out growth scenario, galaxies form stars initially in their central regions.
Consequently, metallicity is expected to decrease with increasing radius from the galaxy center, resulting in a negative metallicity gradient \citep{2014MNRAS.443.3643P}. 
Observational studies have confirmed the existence of such negative metallicity gradient in the Milky Way \citep{stanghelliniGALACTICSTRUCTURECHEMICAL2010, luckDISTRIBUTIONELEMENTSGALACTIC2011}, 
and in other local galaxies \citep[e.g.,][]{sanchezCharacteristicOxygenAbundance2014, hoMetallicityGradientsLocal2015}. 
Furthermore, a correlation between metallicity gradient and stellar mass has been observed, 
where more massive galaxies tend to exhibit steeper negative metallicity gradients \citep{belfioreSDSSIVMaNGA2017, maiolinoReMetallicaCosmic2019}.

However, at higher redshifts, the situation becomes more complex. Studies have revealed heterogeneous metallicity gradient results 
including negative, flat, and even positive (inverted) gradients \citep[e.g.,][]{cresciGasAccretionOrigin2010,queyrelMASSIVMassAssembly2012,swinbankPropertiesStarformingInterstellar2012,jonesORIGINEVOLUTIONMETALLICITY2013,
    2016ApJ...820...84L,2016ApJ...827...74W,molinaSINFONIHiZELSDynamicsMerger2017, wang2017,wang2019,wang2020b,Wang_2022, cartonFirstGasphaseMetallicity2018,schreiberSINSZCSINFSurvey2018,patricioResolvedScalingRelations2019,2020MNRAS.492..821C,simonsCLEARGasphaseMetallicity2021,liFirstCensusGasphase2022,
    venturiGasphaseMetallicityGradients2024,li13BillionYear2025,juMSA3DMetallicityGradients2025,2025arXiv251008997S}.
These diverse metallicity gradients can be explained by the influence from various physical mechanisms. 
For instance, flat metallicity gradients can result from efficient radial mixing caused by galaxy merging/interaction, supernova winds and the outflows of metal enriched gas 
\citep{rupkeGALAXYMERGERSMASS2010,gibsonConstrainingSubgridPhysics2013,floresStarformingRegionsMetallicity2014,maWhyHighredshiftGalaxies2017}.
Conversely, the positive gradients can be attributed to the rapid inflow of pristine, metal-poor gas into the central regions, diluting the core metallicity \citep{cresciGasAccretionOrigin2010, venturiGasphaseMetallicityGradients2024},

To distinguish between these mechanisms, it is essential to investigate galaxies in different environments. While the field population has been
extensively studied, the impact of overdense environments, such as protoclusters, on metallicity gradients remains less explored. Protoclusters at cosmic noon ($z \sim 2-4$) are sites of an accelerated galaxy evolution with
higher merger rates, enhanced gas accretion, potential environmental quenching, and environmentally driven shocks \citep{chiangGalaxyProtoclustersDrivers2017,shimakawaMAHALODeepCluster2018a, zhouMAMMOTHMOSFIREEnvironmentalEffects2025a}. 
Since these processes can significantly influence the metallicity distribution within galaxies, we expect to observe distinct metallicity gradient patterns in protocluster galaxies compared to field galaxies.

An ideal laboratory is provided by the three extremely massive overdense protoclusters BOSS1244, BOSS1441, and BOSS1542 at $z \sim 2.3$, which were discovered by the MApping the Most Massive Overdensity Through Hydrogen (MAMMOTH) project \citep{caiMAPPINGMOSTMASSIVE2016, caiDiscoveryEnormousLya2017,zhengMAMMOTHConfirmationTwo2021, shiSpectroscopicConfirmationTwo2021a}. 

In this work. we present the metallicity gradients of 42 star forming galaxies in these three protoclusters at $z \sim 2.3$ using \textit{Hubble Space Telescope (HST}) grism data. 
In \citet{yangMAMMOTHGrismRevisitingMassMetallicity2025}, we have investigated the stellar mass-metallicity relation (MZR) of galaxies in these three protoclusters, and found a significant metal deficiency in protocluster galaxies compared to the field at a fixed stellar mass. 
In other words, these protoclusters show a strong environmental effect on their member galaxies' integrated metallicity, which motivates us to further explore the impact of environment on the spatial distribution of metallicity within these galaxies.
\citet{liFirstCensusGasphase2022} have studied the metallicity gradients of star forming galaxies in the BOSS1244 protocluster using the same \textit{HST} grism data. 
To ensure consistent measurement methodologies across all three protoclusters and to leverage the larger sample size now available, we do not directly adopt the value previously reported by \citep{liFirstCensusGasphase2022}.
Because different methodologies of estimating stellar mass and metallicity gradients will introduce systematic biases, 
we re-analyze the BOSS1244 protocluster data using our current framework. 

Our paper is organized as follows. In Section~\ref{sec:obs}, we summarize the observations and data reduction procedure.
In Section~\ref{sec:methods}, we describe our sample selection, Voronoi binning method, and metallicity gradient measurement.
Then we present our main results in Section~\ref{sec:results}.
Finally, we summarize our conclusions in Section~\ref{sec:conclusion}. 
Throughout this paper, we adopt the AB magnitude system and the standard concordance cosmology ($\Om=0.3, \Ol=0.7$, $H_0=70\,\Hunit$).  
The metallic lines are denoted in the following manner, if presented without wavelength: 
$\OIII\lambda$5008$\defeq$\OIII,
$\OII\lambda\lambda$3727,3730$\defeq$\OII,
$\NeIII \lambda $3869$\defeq$\NeIII, and
$\NII{\lambda}$6585$\defeq$\NII.

\section{Observation and Data Reduction} \label{sec:obs}
We used the same spectroscopic data as \citet{yangMAMMOTHGrismRevisitingMassMetallicity2025}, hereafter Y26, 
for the three protoclusters at $z \sim 2.3 $, obtained through 
the MAMMOTH-Grism survey, an \textit{HST} Cycle 28 medium program (GO: 16276, P.I.: X. Wang). This program utilized the Hubble Space Telescope's 
Wide Field Camera 3 (WFC3) with the F125W filter and G141 grism to perform deep near-infrared imaging and spectroscopy, targeting the central 
regions of the most massive galaxy protoclusters with a total allocation of 45 primary orbits. We also leveraged the existing multi-wavelength 
imaging data in these regions, including \textit{HST/WFC3} F160W band, LBT/LBC Uspec, V-BESSEL, and z-SLOAN bands, CFHT/WIRCam Ks band, and KPNO-4m/MOSAIC Bw band imaging data. The exact photometric coverage varies among the different protocluster fields. Specifically, for BOSS1244 and BOSS1542, the available photometry includes the F125W, F160W, Uspec, z-SLOAN, and Ks bands. For BOSS1441, the available photometry includes the F125W, F160W, Uspec, V-BESSEL, and Bw bands. Therefore, each galaxy in our sample is consistently fitted using 5 photometric data points.
 
The details of these data and the data reduction process were described in Y26. Here, we briefly summarize the main steps of the data reduction process.

We first utilized the \texttt{GRIZLI}\footnote{\url{https://github.com/gbrammer/grizli/}} software \citep{Grizli} to reduce the G141 data, which yielded the following 
results: (1) the 1D and 2D grism spectra of each source, and (2) the 2D emission-line 
maps extracted from individual extended sources.
Based on the extracted 1D grism spectra, we then measured the redshifts following the method described in Appendix A of \citet{wang2019}. 
Simultaneously, we fit the intrinsic nebular emission line fluxes using 1D Gaussian profiles centered at their respective wavelengths. 
As a result. we obtained the fluxes of \OII, \NeIII, \Hd, \Hg, \Hb, and \OIII for each source.

There are some resolution differences between our ground-based
and space-based imaging data. 
To fix this, we first utilized the \texttt{Source Extractor} \citep{1996A&AS..117..393B} software in dual mode on space-based data (F160W and F125W imaging) 
to extract source models and the fluxes.
Then we used the \texttt{T-PHOT} \citep{merlinTPHOTVersion202016} software on ground-based data using the models obtained from the more detailed space-based imaging data, 
which allows us to obtain compatible accurate photometric measurements. In conclusion, 
we obtain broadband photometry covering the wavelength range of [1000, 7000] \AA\ in the rest frame for sources in the three protoclusters at $z \sim 2.3$.
As a result, each source of the three protoclusters has 5 photometric data points.

\section{Methods} \label{sec:methods}

\subsection{Sample Selection} \label{sec: Selection}
In Y26, we have selected a sample of 63 star forming galaxies in three protoclusters at $z \sim 2.3$.
The selection procedure is summarized as follows.
\begin{itemize}
    \item We first selected galaxies with secure grism redshifts ranging from 2.15 to 2.35, 
    which is the redshift range of the three protoclusters.
    \item We then selected galaxies with their \OIII\ and \OII\ emission lines securely detected 
    with S/N $>$ 3, and we only included the galaxies without obvious contamination in their 1D spectra.
    \item Finally, we excluded active galactic nuclei (AGNs) using the MEx diagram proposed by \cite{2011ApJ...736..104J, 2014ApJ...788...88J} and modified by \cite{2015ApJ...801...35C}.
\end{itemize}

We further selected our sample galaxies in this paper through careful visual inspection of their morphologies in high-resolution \textit{HST} imaging. This step ensures that the galaxies show well-defined, extended structures suitable for spatially resolved analysis, and excludes systems undergoing obvious mergers. After this initial cut, we imposed a strictly quantitative requirement regarding the spatial extent of each galaxy. Specifically, a galaxy is only included in our final sample if the Voronoi binning (VorBin) technique (see Section~\ref{sec:Vorbin}) yields a minimum of 5 valid bins. This minimum number is essential to reliably constrain a radial metallicity gradient.

\begin{figure*}[htbp!]
    \centering
    \includegraphics[width=0.9\textwidth]{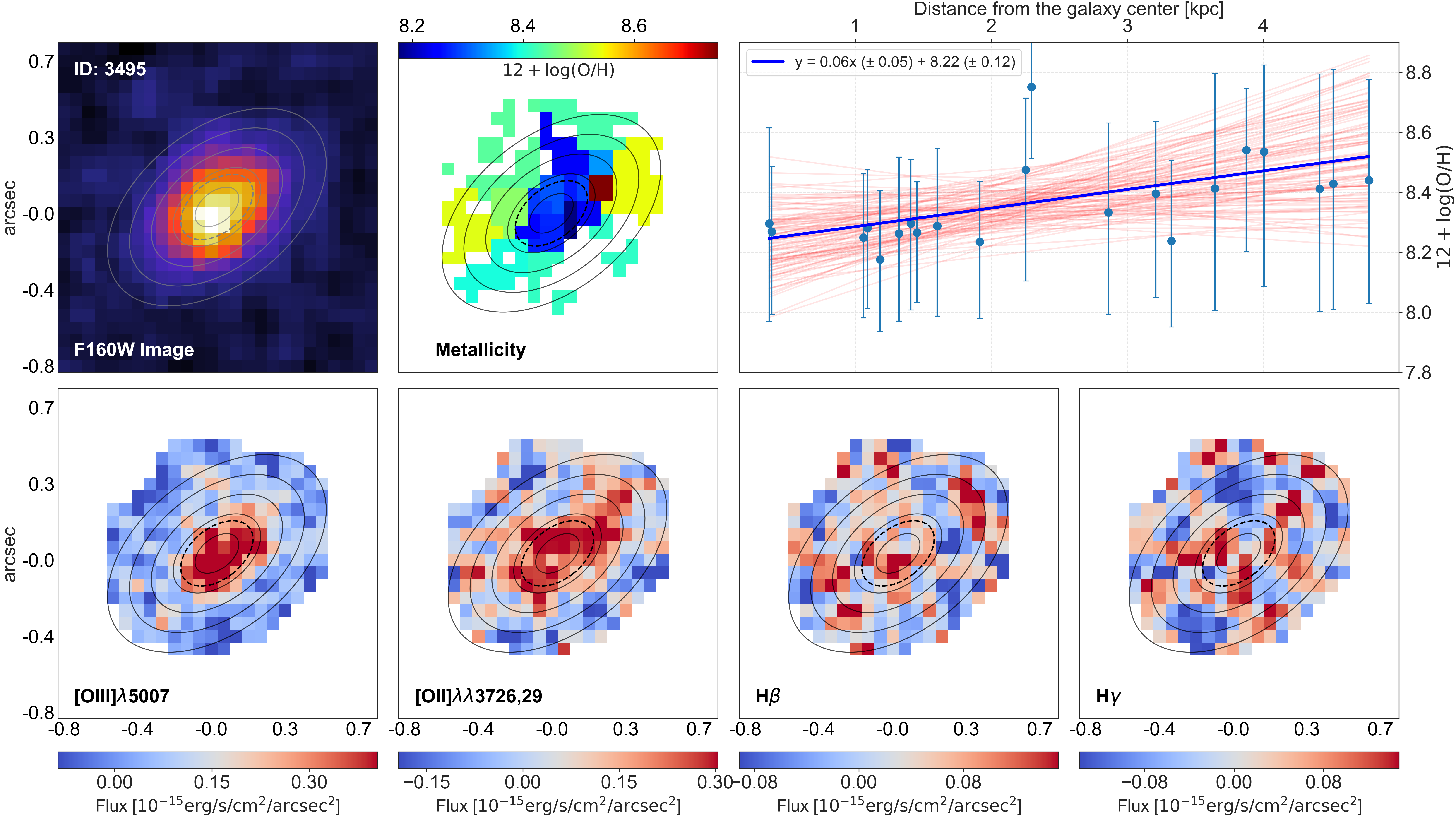}
    \caption{Example of spatially resolved metallicity measurements of the protocluster member galaxy ID3495 at $z\sim2.3$. 
    The top row, from left to right, shows the F160W image cutout, Voronoi binned 2D metallicity map, and the linear fit to the metallicity gradient.
    The lower row displays the \OIII, \OII, \Hb, \Hg\ emission-line maps from left to right. In the F160W image cutout, the metallicity map, and the emission-line maps, the solid gray ellipses mark galactocentric radii of 1, 2, 3, 4, and 5 kpc, while the dotted gray ellipse denotes the effective radius of the galaxy. In the metallicity-gradient panel, the solid blue line indicates the best-fit linear relation, and the blue error bars represent the 1$\sigma$ measurement uncertainties of the each bin. The thin red lines visualize 100 random draws from the linear regression 
    }
    \label{fig:voronoi_example}
\end{figure*}

\subsection{Properties Derived in Our Previous Work}

The stellar mass (\Mstar) and metallicity ($Z$) of the sample galaxies have 
been obtained in Y26. We briefly summarize the methods used to derive these properties below.

We derived the \Mstar\ by fitting the multi-wavelength photometry using the \texttt{BAGPIPES} software \citep{2018MNRAS.480.4379C} with the following assumptions:
(1) the \citet{2003MNRAS.344.1000B} (BC03) stellar population synthesis models combined with a \citet{2003PASP..115..763C} initial mass function (IMF), (2) an extinction law from \citet{Av2000C}, (3) an exponentially declining star formation history (SFH), and (4) a fixed redshift from the grism spectra. Specifically, 
in the \texttt{BAGPIPES} SED fitting, the stellar metallicity and dust attenuation were treated as free parameters, independent of the nebular properties derived from the emission lines. We adopted uniform priors for the stellar metallicity ranging from 0.02 to 2.5 $Z_\odot$, and for the dust attenuation $A_V$ ranging from 0 to 3 magnitudes.

The metallicity was derived by fitting the observed emission-line fluxes (\OII , \OIII, \Hg, \Hb)
 using the forward-modeling Bayesian inference method described and used in \citet{wang2017,wang2019,wang2020b,Wang_2022} and Y26, which is superior to conventional methods because, instead of using line ratios, this method forward-models all available emission-line fluxes directly, maintaining consistency with faint, low-SNR \Hb lines \citep[see][Figure 6]{wang2020b}. 
 The details of the fitting procedure are described in Section \ref{sec:MetallicityGradient}.

\begin{figure*}[htbp!]
    \centering
    \includegraphics[width=0.9\textwidth]{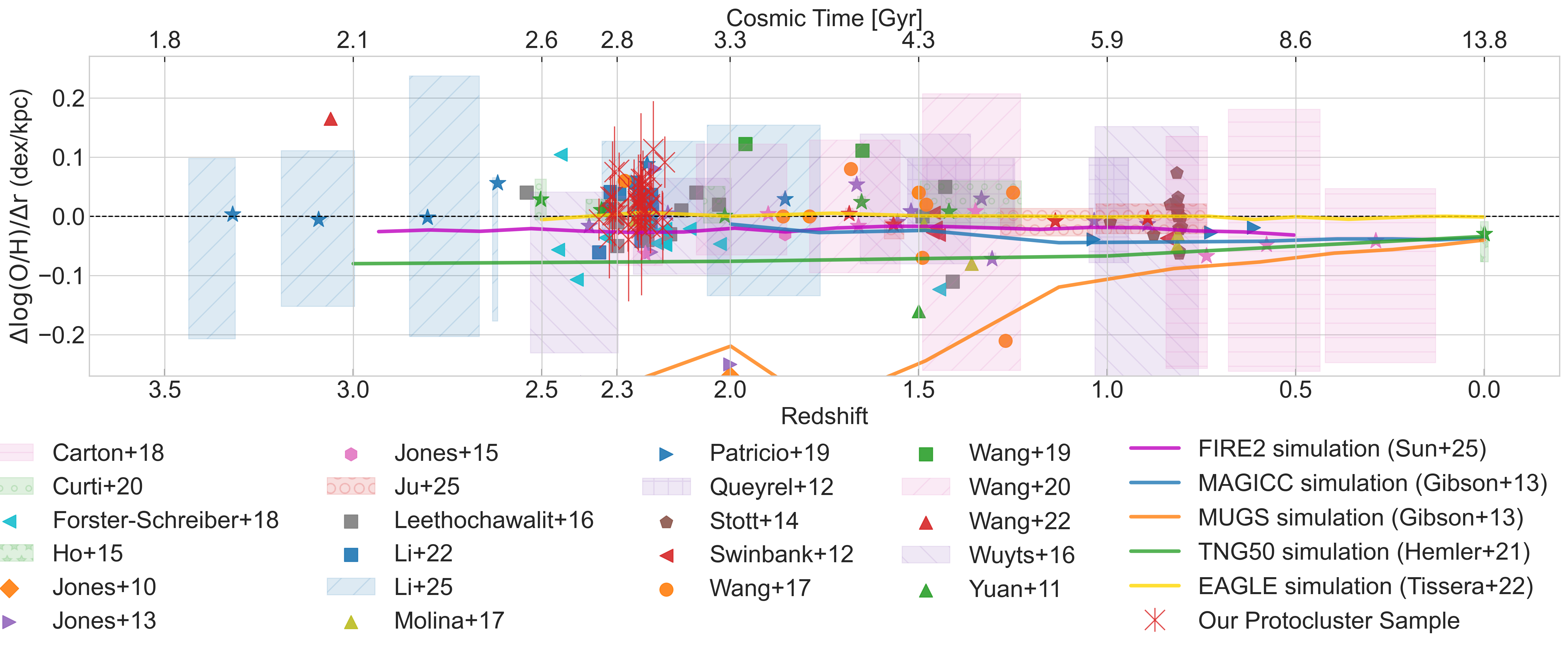}
    \caption{
            Evolution of metallicity gradients (in dex kp$\rm c^{-1}$) as a function of redshift and cosmic time. 
            We present measurements for our sample of $N=42$ protocluster members at $z\sim2.3$, shown as red 'x' symbols with $1 \sigma$ error bars. 
            The horizontal dashed line indicates a flat gradient ($\nabla_r Z = 0$). 
            Literature observations are plotted as various colored symbols; for studies with large samples ($N > 20$), we show the data range as shaded boxes and the median value as a star symbol. 
            These comparison samples include 
            \citet{jonesMEASUREMENTMETALLICITYGRADIENT2010}, 
            \citet{yuanSystematicsMetallicityGradient2013}
            \citet[GLASS]{jonesORIGINEVOLUTIONMETALLICITY2013,jonesGRISMLENSAMPLIFIEDSURVEY2015}, 
            \citet[HiZELS]{swinbankPropertiesStarformingInterstellar2012}, 
            \citet[MASSIV]{queyrelMASSIVMassAssembly2012}, 
            \citet{hoMetallicityGradientsLocal2015}
            \citet[CASSOWARY]{2016ApJ...820...84L}, 
            \citet[KMOS3D]{2016ApJ...827...74W}, 
            \citet{molinaSINFONIHiZELSDynamicsMerger2017}, 
            \citet[GLASS + GLASS JWST]{wang2017,wang2019,wang2020b,Wang_2022}, 
            \citet{cartonFirstGasphaseMetallicity2018}, 
            \citet[SINS/zC-SINF]{schreiberSINSZCSINFSurvey2018}, 
            \citet{patricioResolvedScalingRelations2019}, 
            \citet[KLEVER]{2020MNRAS.492..821C}, 
            \citet{liFirstCensusGasphase2022, li13BillionYear2025}, 
            \citet{venturiGasphaseMetallicityGradients2024}, 
            and 
            \citet{juMSA3DMetallicityGradients2025}. 
            Solid colored lines represent predictions from cosmological simulations, specifically 
            MUGS and MAGICC \citep{gibsonConstrainingSubgridPhysics2013}, 
            FIRE2 \citep{sunPhysicalOriginPositive2025}, 
            TNG50 \citep{hemlerGasphaseMetallicityGradients2021}, and 
            EAGLE \citep{tisseraEvolutionOxygenAbundance2022}. 
            While our data overlap with the range of previous measurements, the protocluster sample tends to favor flatter/positive gradients compared with field measurements.
            }
    
    \label{fig:MGZ}
\end{figure*}

\subsection{Voronoi Binning} \label{sec:Vorbin}
We applied the adaptive Voronoi binning algorithm \citep{2003MNRAS.342..345C} 
with the \texttt{vorbin} package to the emission-line maps. 
The package uses a centroidal Voronoi tessellation algorithm to group pixels into spatial bins based on a specified S/N threshold.
In our analysis, we set the target S/N for each bin to 3.5, 
set by the brightest available line (\OIII), which preserves resolution in high-SNR
regions while retaining information in the outskirts. The resulting tessellations follow the galaxy morphology 
and provide the basis for measuring metallicity gradients in the next section. The emission-line flux in each bin is obtained by summing the fluxes of all pixels within that bin.
The example figure of Voronoi binning result is shown in the upper middle panel of Figure~\ref{fig:voronoi_example}, which is also the metallicity map. 
We can see some blank regions in the Voronoi bin map, which are the areas where the S/N of \OIII\ is too low to reach the target S/N even after binning are therefore masked out. As a result, for valid bins in our whole sample, the median S/N and the 16th-84th percentile ranges for all relevant lines are: \OIII\ (4.95, [3.88, 7.43]), \OII\ (1.68, [1.02, 2.35]), \Hb\ (1.22, [0.28, 2.24]), and \Hg\ (0.25, [0.00, 1.18]). 


\subsection{Metallicity Gradient Measurement} 
\label{sec:MetallicityGradient}
We estimated the metallicity in each spatial bin following the forward-modeling Bayesian inference method used to estimate the integrated metallicity in Y26. Specifically, we adopted the \citet{2018ApJ...859..175B} calibration, utilizing the BC03 stellar population synthesis models with a \citet{2003PASP..115..763C} initial mass function.
We utilized the  fluxes and the uncertainties of \OII, \OIII, \Hb, and \Hg in each bin to derive the metallicity ($Z$), nebular dust extinction ($A_V^{\rm}$) and dereddened \Hb\ flux ($f_{\rm H\beta}$).

The likelihood is defined as:

\begin{equation}
    {
    L \propto \exp\left(-\frac{1}{2} \sum_{i=1}^{N_{\rm lines}}\frac{f_{EL_i}-R_i\cdot f^2_{\rm H\beta}}{\sigma ^2_{EL_i} + f^2_{\rm H\beta} \cdot \sigma ^2_{R_i}}\right)
    }
\end{equation}
\begin{displaymath}
\ 
\end{displaymath}
where $R_{i}$ is the theoretical line flux ratio, such as the Balmer decrement ($\mathrm{H}\beta/\mathrm{H}\gamma = 0.47$) or a metallicity-sensitive ratio like $\OIII/\mathrm{H}\beta$. The measured emission-line fluxes and their uncertainties are
 $f_{\mathrm{EL}}$ and $\sigma_{\mathrm{EL}}$. Posterior distributions for the model parameters were obtained using the \texttt{emcee} MCMC sampler \citep{emcee}.

The de-projected distance from the geometric center of each bin to the galaxy center (in kpc) was derived using structural parameters measured with \texttt{GALFIT} \citep{2002AJ....124..266P}, including the galaxy center, axis ratio, and position angle.  All of our sample galaxies were fitted with 2D S\'ersic profiles to the direct F160W image. It should be noted that no attempt is made to account for the finite spatial resolution of the observations when measuring the metallicity gradients. For our sample, the typical PSF-to-galaxy-size (effective radius) ratio has a median value of 0.66.
We then perform a weighted linear regression on the measured metallicities and de-projected distances with measurement errors taken into account. The weights were determined using the 1$\sigma$ uncertainty of each data point,  The best-fit slope of this linear relation is defined as the metallicity gradient, and the uncertainty of the slope is derived from the covariance matrix of the fit. An example of the metallicity gradient measurement is shown in the upper right panel of Figure~\ref{fig:voronoi_example}. In this panel, the blue points with error bars represent the metallicity measurements in each Voronoi bin, the solid blue line indicates the best-fit linear relation, and we also plot 100 randomly selected lines draws from the linear regression.

\section{Results} \label{sec:results}
Our final sample consists of 42 star forming galaxies in three protoclusters at $z \sim 2.3$ with reliable metallicity gradient measurements.
Among them, 29 galaxies ($\sim69\%$) exhibit flat or positive metallicity gradients (defined as the best estimate of the gradient $ > 0$), while 13 galaxies show negative gradients. 

It is worth reminding readers that,  as noted by \citet{yuanSystematicsMetallicityGradient2013}, superficially flat metallicity gradients can be partially caused by limited angular resolution. However, the metallicity maps in our sample often show complex, non-radially symmetric patterns, indicating that at least part of the observed diversity reflects intrinsic, non-axisymmetric metal distributions rather than resolution effects alone.
\begin{figure*}[htbp!]
    \centering
    \includegraphics[width=0.9\textwidth]{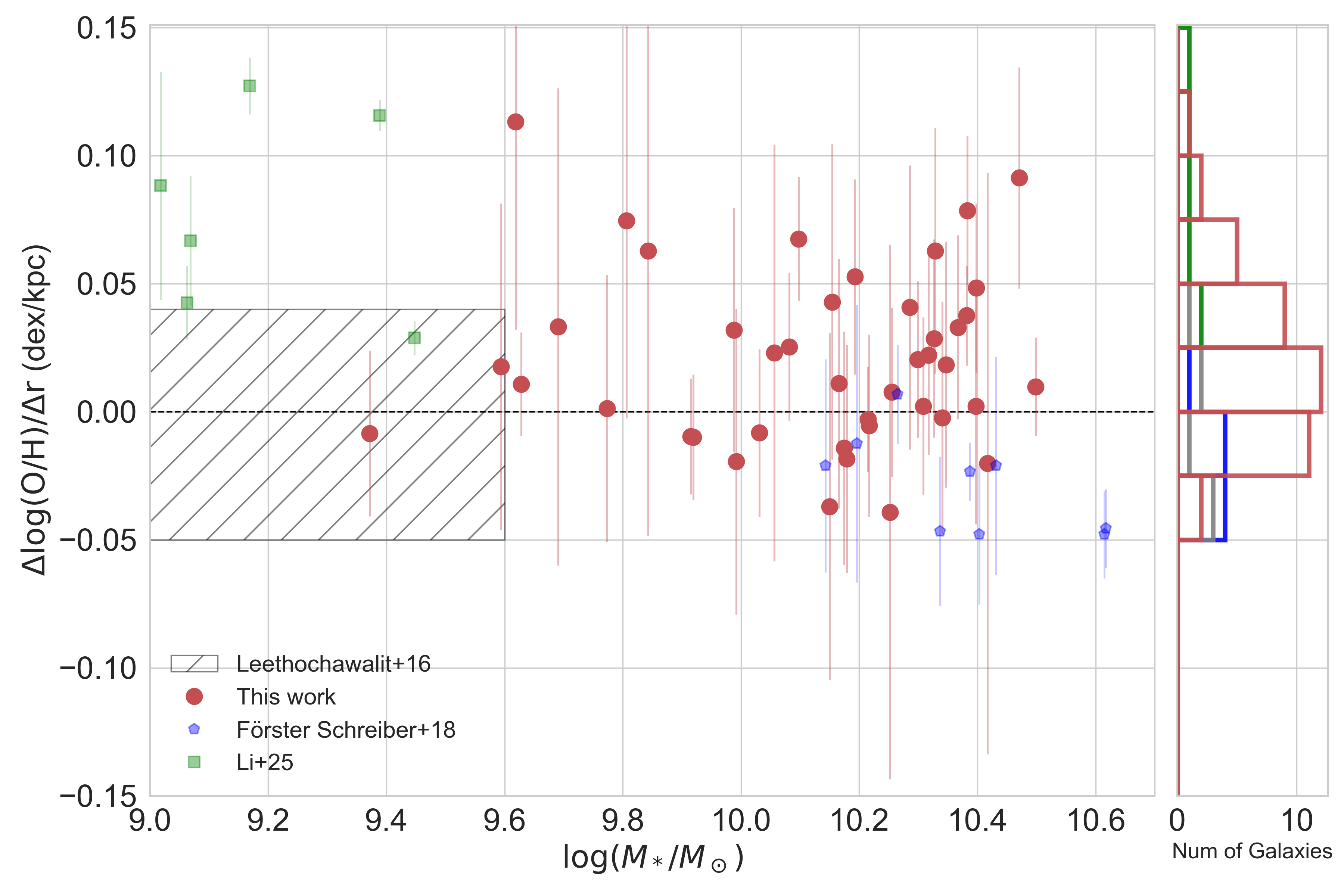}
    \caption{
        \textbf{Left:} Metallicity gradient as a function of stellar mass. 
        Our protocluster measurements at $z\sim2.3$ are shown as red dots, with red contours indicating the number density of the sample. 
        The horizontal dashed line marks a flat gradient ($\nabla_r Z = 0$).
        We compare against field galaxy samples ($1.7<z<2.4$) shown as colored symbols 
        \citep{schreiberSINSZCSINFSurvey2018, li13BillionYear2025} 
        and the hatched gray region representing the range from \citet{2016ApJ...820...84L}.
        Error bars denote $1\sigma$ uncertainties on the gradients. 
        \textbf{Right:} Marginalized histogram of the metallicity gradient distributions, with colors corresponding to the samples in the left panel.
        The protocluster sample displays a systematic excess of flat and positive gradients relative to the co-eval field population, particularly at the high-mass end ($\log(M_*/M_\odot) \gtrsim 10.0$).
    }
    \label{fig:MG-mass}
\end{figure*}

\begin{figure*}[htbp!]
    \centering
    \includegraphics[width=0.9\textwidth]{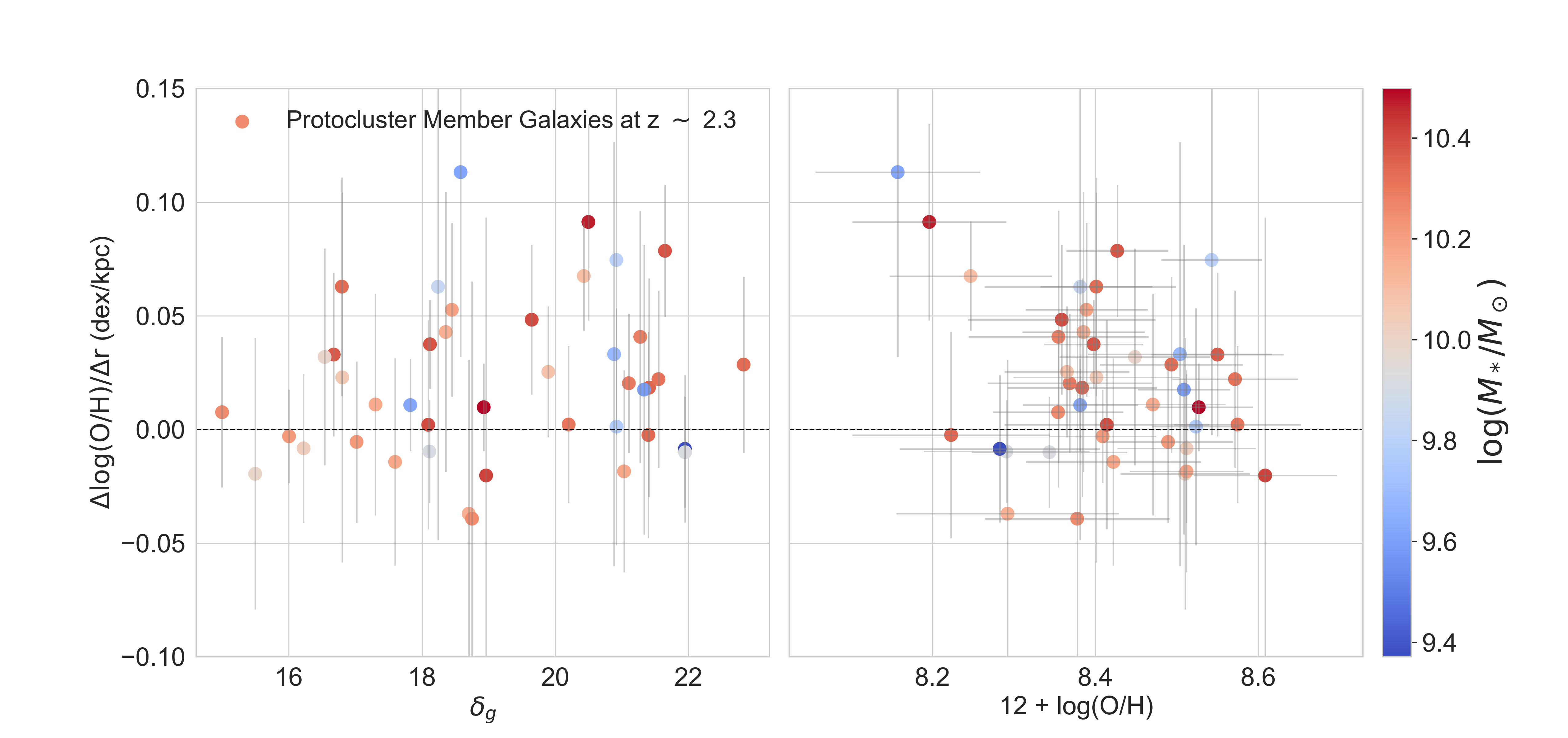}
    \caption{
        \textbf{Left:} Metallicity gradient as a function of local galaxy overdensity $\delta_g$. 
        \textbf{Right:} Metallicity gradient as a function of integrated gas-phase metallicity ($12+\log(\rm O/H)$). 
        In both panels, data points are color-coded by stellar mass ($\log(M_*/M_\odot)$), and error bars represent $1\sigma$ measurement uncertainties. 
        The horizontal dashed line indicates a flat gradient ($\nabla_r Z = 0$). 
        We do not observe a clear dependence of the gradient on either local environment or global metallicity; Spearman correlation analysis yields no statistically significant trends ($p > 0.15$) for either relation.
    }
    \label{fig:DgAndMe}
\end{figure*}

\begin{figure*}[htbp!]
    \centering
    \includegraphics[width=0.9\textwidth]{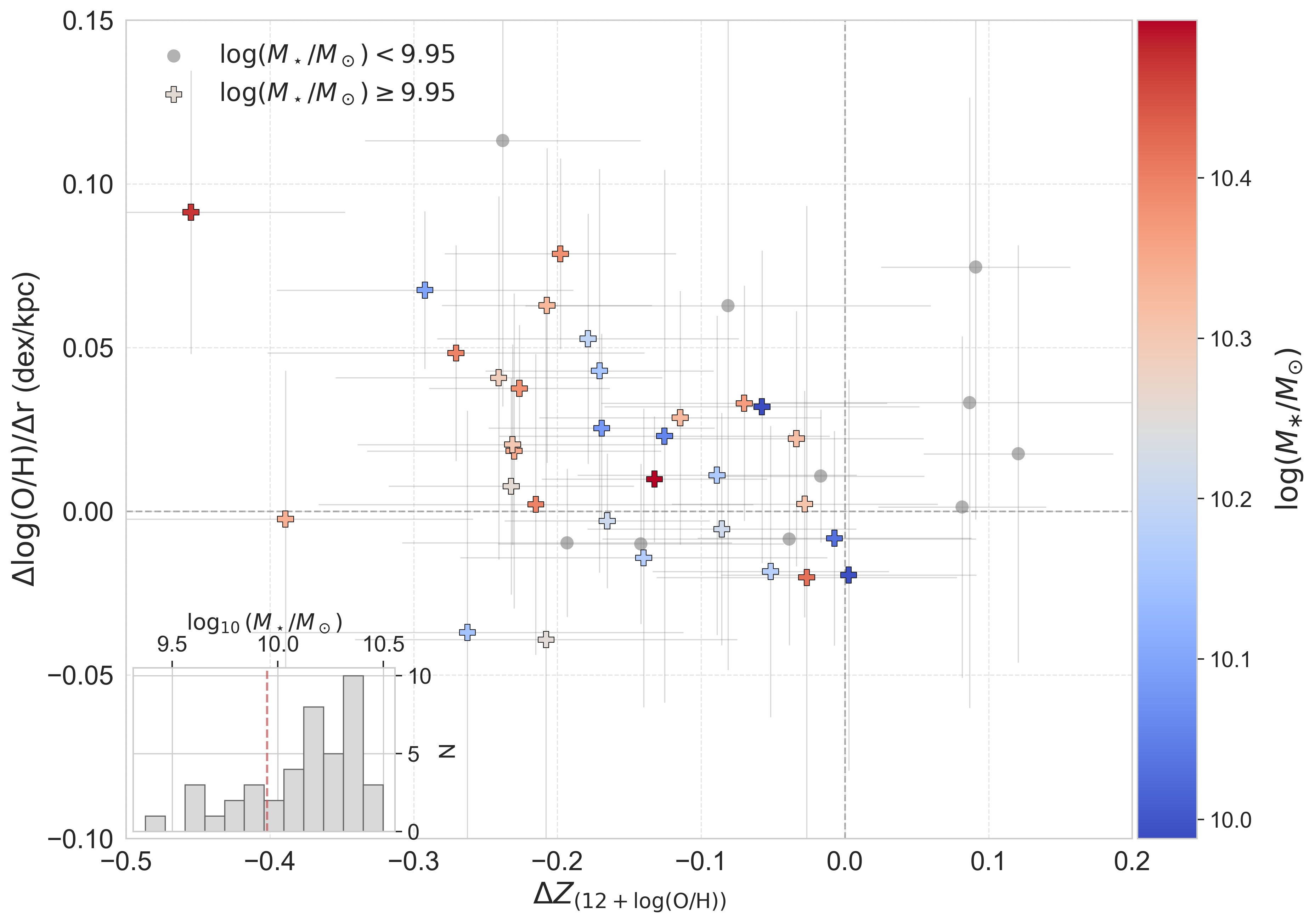}
    \caption{
        Metallicity gradient ($\Delta\log({\rm O/H})/\Delta r$) as a function of the metallicity offset ($\Delta Z$) from the field mass-metallicity relation at $z\sim2.3$. 
        To highlight the mass dependence, the sample is divided by a stellar mass threshold of $\log(M_*/M_\odot) = 9.95$. 
        Galaxies above this threshold are color-coded by stellar mass, while lower-mass galaxies are shown as gray points.
        The inset histogram displays the stellar mass distribution of the sample, with the vertical red dashed line marking the threshold.
        Vertical and horizontal dashed lines mark zero offset and zero gradient, respectively. 
        The massive subsample (colored cross markers) exhibits a strong negative correlation, indicating that massive galaxies which are more metal-poor relative to the MZR are more likely to exhibit positive (inverted) gradients, consistent with a scenario of central chemical dilution driven by the inflow of metal-poor gas.
    }
    \label{fig:Offset}
\end{figure*}

\subsection{Metallicity Gradient versus Redshift}
To place our measurements in a broader evolutionary context, we compare the metallicity gradients of our protocluster members at $z\sim2.3$ with literature observations 
and simulation predictions spanning $0 < z \lesssim 3.5$ in Figure~\ref{fig:MGZ}. The comparison sample comprises a compilation of observational results from the other studies 
\citep{jonesMEASUREMENTMETALLICITYGRADIENT2010,queyrelMASSIVMassAssembly2012,swinbankPropertiesStarformingInterstellar2012,yuanSystematicsMetallicityGradient2013, jonesORIGINEVOLUTIONMETALLICITY2013,jonesGRISMLENSAMPLIFIEDSURVEY2015,2016ApJ...820...84L,2016ApJ...827...74W,molinaSINFONIHiZELSDynamicsMerger2017, wang2017,wang2019,wang2020b,Wang_2022, cartonFirstGasphaseMetallicity2018,schreiberSINSZCSINFSurvey2018,patricioResolvedScalingRelations2019,2020MNRAS.492..821C,liFirstCensusGasphase2022, venturiGasphaseMetallicityGradients2024,li13BillionYear2025,juMSA3DMetallicityGradients2025}.
Our protocluster galaxies occupy the relatively more positive end of the gradient distribution compared to field samples at similar redshifts. While we do not observe a clear evolution trend within our own sample, likely due to the limited redshift distribution of our protoclusters,  the systematic offset toward flatter and inverted gradients is distinct. 
When comparing with simulations \citep[e.g.,][]{gibsonConstrainingSubgridPhysics2013,hemlerGasphaseMetallicityGradients2021,tisseraEvolutionOxygenAbundance2022,sunPhysicalOriginPositive2025}, 
we find that current hydrodynamical simulations generally struggle to reproduce the extreme positive gradients observed in our sample. While simulations with strong feedback (such as FIRE2) predict flatter gradients at high redshift, 
they rarely reach the highly inverted values seen in these protocluster members. This discrepancy suggests that the dense protocluster environment may drive mechanisms (such as rapid, metal-poor gas inflows and merger-induced mixing) that are not fully captured in these simulations.

\subsection{Metallicity Gradient versus Stellar Mass} \label{sec:mass}
We present the distribution of metallicity gradients and stellar masses for our protocluster sample in the left panel of Figure~\ref{fig:MG-mass} in red dots with $1 \sigma$ error bars. 
For comparison, we include a compiled sample of field galaxies in the redshift range $1.7 < z < 2.4$. 
To ensure a fair comparison and minimize the heterogeneity introduced by varying spatial resolutions, we restrict this field sample to studies utilizing space-based observations \citep{li13BillionYear2025} or AO-assisted ground-based observations\citep{2016ApJ...820...84L, schreiberSINSZCSINFSurvey2018}. Furthermore, similar to our analysis, these reference studies do not attempt to account for finite spatial resolution effects when deriving their gradients.
As shown in the right panel of Figure~\ref{fig:MG-mass}, the protocluster sample spans a range of gradients with a median value that is notably more positive than field galaxy samples at similar redshifts. To statistically quantify this difference, we performed a mass-matched Mann-Whitney U test between the two datasets. The test yields a p-value of $p = 0.0028$ ($2.77\sigma$).
When taking stellar mass into account, we find that the protocluster galaxies exhibit a significantly higher fraction of positive metallicity gradients compared to the field sample at fixed stellar mass, particularly among massive galaxies with $\log(M_*/M_\odot) > 10.0$. 
Overall, we do not observe a clear global trend between stellar mass and metallicity gradient in our protocluster sample. 
This is in contrast to the negative correlation reported in some field galaxy studies and simulations \citep[e.g.,][]{belfioreSDSSIVMaNGA2017, maiolinoReMetallicaCosmic2019, juMSA3DMetallicityGradients2025, hemlerGasphaseMetallicityGradients2021, sunPhysicalOriginPositive2025}. 

\subsection{Metallicity Gradient versus Local Density and Integrated Metallicity} \label{sec:Met}
To investigate potential environmental drivers of the observed gradients, we examine the relationship between metallicity gradients, local overdensity, and the integrated gas-phase metallicity of the galaxies.
We quantify the local environment using the galaxy overdensity parameter, $\delta_g$, defined as $\delta_g = (\Sigma_{\rm group}/\Sigma_{\rm field}) - 1$, where $\Sigma_{\rm group}$ and $\Sigma_{\rm field}$ denote the surface densities of H$\alpha$ or Ly$\alpha$ emitters within the protocluster and in random fields at similar redshifts, respectively \citep{caiDiscoveryEnormousLya2017,shiSpectroscopicConfirmationTwo2021a,yangMAMMOTHGrismRevisitingMassMetallicity2025}.
Specifically, we adopt the local $\delta_g$ values derived from the density maps constructed by Y26, which employed a 2D Gaussian kernel density estimation (KDE) with a smoothing scale of $150''$ ($\sim 4.0$ cMpc).
Figure~\ref{fig:DgAndMe} presents the metallicity gradient as a function of local overdensity $\delta_g$ (left panel) and integrated oxygen abundance $12+\log(\rm O/H)$ (right panel). The data points are color-coded by stellar mass. 
As shown in the left panel, we find no strong or statistically significant correlation between the measured gradients and the local density $\delta_g$. This suggests that, within the protocluster structure, local density variations on $\sim 4$ cMpc scales may not be the primary factor governing metal distribution.
Similarly, we observe no clear global trend between the metallicity gradient and the integrated metallicity (right panel) across the full sample. Spearman correlation analysis yields no statistically significant trends ($p > 0.15$) for these relations. However, if we restrict our analysis to the massive galaxy population ($\log(M_*/M_\odot) > 10.0$), a tentative negative trend emerges, where metal-rich galaxies tend to exhibit more negative gradients. We also explore the implications of this mass-dependent behavior further in Section~\ref{Massdiff}.

\subsection{Metallicity Gradient versus Integrated Metallicity Offset} \label{Massdiff}
We further investigate the connection between the metallicity gradient and the global metal budget by comparing our galaxies to the expected MZR of field galaxies at this redshift. 
We compute the metallicity offset $\Delta Z$ for each protocluster galaxy relative to the field MZR derived from the MOSDEF survey \citep{sandersMOSDEFSurveyEvolution2021}. We utilize the \citet{sandersMOSDEFSurveyEvolution2021} relation specifically because it used the same metallicity calibration (B18) as our work, ensuring a consistent baseline.

As discussed in Sections \ref{sec:mass} and \ref{sec:Met}, we observe no strong global trends between metallicity gradient and either stellar mass or integrated metallicity across the full protocluster sample. 
However, tentative trends emerge when the analysis is restricted to the massive galaxy population ($\log(M_*/M_\odot) > 10.0$).
To rigorously quantify this mass-dependent behavior, we examine the correlation between the metallicity gradient and the integrated metallicity offset ($\Delta Z$) within cumulative subsamples defined by increasing stellar mass thresholds, ranging from $\log(M_*/M_\odot) = 9.6$ to $10.35$.
We observed an evolution: the anti-correlation between the metallicity gradient and the metallicity offset strengthens and becomes more statistically significant as the mass threshold increases.

The correlation strength peaks with a Pearson correlation coefficient of $r = -0.43$ and high significance ($p = 0.015$) for the subsample restricted to massive galaxies with the threshold $\log(M_*/M_\odot) = 9.95$ ($N=32$).
To illustrate this distinction, we present the relation in Figure \ref{fig:Offset} with a visual separation: galaxies below this mass threshold are shown in gray, while the massive population $\log(M_*/M_\odot) = 9.95$ is color-coded by stellar mass.
Using this threshold, we also examined the correlation between the metallicity gradient and other factors.
We found that the metallicity gradient is also anti-correlated with the integrated metallicity with a slightly lower significance compared with that of offsets ($r = -0.354$, $p = 0.047$).
In contrast, the metallicity gradient shows no statistically significant dependence on either local overdensity ($r = 0.27$, $p = 0.136$) or stellar mass itself ($r = 0.26$, $p = 0.157$).

These results suggest that for massive protocluster galaxies, the integrated metallicity offset (or integrated metallicity) is the dominant factor of the gradient, rather than the local environment or stellar mass.
This anti-correlation aligns with the "pristine-gas inflow" scenario proposed by \citet{cresciGasAccretionOrigin2010} (and see a graph in \citet[Fig. 1]{venturiGasphaseMetallicityGradients2024}) to explain positive gradients in three Lyman-break galaxies at $z\sim3$. Physically, the rapid infall of pristine (metal-poor) gas toward the galaxy center dilutes the central metallicity while leaving the outskirts relatively enriched, simultaneously lowering the global metallicity (negative $\Delta Z$) and inverting the gradient (positive slope).
Crucially, our sample is drawn from dense protocluster environments where cold gas accretion is predicted to be significantly enhanced compared to the field \citep{wangEnvironmentalDependenceMass2023, daikuharaAssociationColdGas2025a}. This environmental enhancement of inflows provides a natural explanation for the excess of positive gradients and the observed correlation with metal deficiency in our sample. Given that our interpretation invokes enhanced gas accretion, it is relevant to examine the star formation activity of our sample. We investigated the star formation rates (SFRs) of our protocluster galaxies, derived from the dereddened \Hb\ flux as detailed in Y26, and compared them to the field star-forming main sequence at comparable redshifts. We find that our protocluster galaxies are generally consistent with the main sequence and do not exhibit systematically higher specific SFRs at a fixed mass. This lack of starburst signature could be driven by selection effects, as our emission-line selected sample could miss heavily dust-obscured starburst populations. What's more, the accreted pristine gas might possess high angular momentum or is temporarily heated before settling into the central star-forming regions.


\section{Conclusion} \label{sec:conclusion}
In this work, we present measurements of gas-phase metallicity gradients for a sample of 42 star-forming galaxies in three protoclusters (BOSS1244, BOSS1441, and BOSS1542) at $z\sim2.3$, utilizing HST slitless grism spectroscopy from the MAMMOTH-Grism program.
A key finding is that a significant fraction (69\%) of these protocluster members exhibit flat or positive (inverted) metallicity gradients, which is in contrast to the flat or negative gradients observed in field galaxies at similar redshifts.

We further investigated the dependence of metallicity gradients on stellar mass, local galaxy overdensity, integrated metallicity, and the integrated metallicity offset from the field MZR. The principal results are summarized as follows:

\begin{itemize}
    \item We observe no strong trends between the metallicity gradient and either local galaxy overdensity or stellar mass for the full protocluster sample. This suggests that the gradient is not simply determined by the local density $\delta_g$ or stellar mass.
    
    \item Similarly, we find no significant correlation between the metallicity gradient and integrated metallicity for the full protocluster sample. However, when the analysis is restricted to massive galaxies ($\log(M_*/M_\odot) > 10.0$), a tentative negative trend emerges, where metal-rich galaxies tend to exhibit more negative (steep) gradients.
    
    \item We find a moderate, statistically significant negative correlation between the metallicity gradient and the integrated metallicity offset ($\Delta Z$) relative to the field MZR. This indicates that galaxies which are metal-poor for their mass are more likely to exhibit positive (inverted) gradients. This result supports a scenario of "pristine-gas inflow" driven by the rapid inflow of pristine, metal-poor gas.
    
    \item By analyzing the gradient-offset ($\Delta Z$) correlation within cumulative subsamples, we found that this anti-correlation is mass-dependent. The relationship strengthens and increases in statistical significance as the mass threshold is raised, peaking at the threshold $\log(M_*/M_\odot) = 9.95$ ($r \approx -0.43$, $p = 0.015$).
\end{itemize}

The correlation between metallicity gradients and metal deficiency, particularly among massive protocluster galaxies, suggests that the environment-enhanced inflow of metal-poor gas plays a crucial role in shaping the internal chemical distribution of these systems. 
Conversely, the absence of such trends in the low-mass regime may be due to effective recycling of metal-enriched feedback-driven winds/outflows in overdense environments \citep{oppenheimerMassMetalEnergy2008}, which is more pronounced in low-mass galaxies \citep{el-badryGasKinematicsMorphology2018}.

Future high-resolution spectroscopic observations with \textit{JWST} or ground-based AO-assisted IFU instruments will be essential to spatially resolve the gas kinematics and further elucidate the physical mechanisms driving these trends. Additionally, testing these correlations with larger statistical samples of both protocluster and field galaxies at different redshifts is necessary to confirm the environmental dependence of metallicity gradients and their connection to cold gas accretion processes.


\begin{acknowledgments}
We thank the anonymous referee for a careful read and useful comments that helped improve the clarity of this paper. 
This work is supported by the National Key R\&D Program of China No.2025YFF0510603, the National Natural Science Foundation of China (grant 12373009), the CAS Project for Young Scientists in Basic Research Grant No. YSBR-062, the China Manned Space Program with grant no. CMS-CSST-2025-A06, and the Fundamental Research Funds for the Central Universities. XW acknowledges the support by the Xiaomi Young Talents Program, and the work carried out, in part, at the Swinburne University of Technology, sponsored by the ACAMAR visiting fellowship. CWT is supported by NSFC 12588202. This work is also supported by NASA through \textit{HST} grant HST-GO-16276 and HST-GO-17159.
The \textit{HST} data presented in this article were obtained from the Mikulski Archive for Space Telescopes (MAST) at the Space Telescope Science Institute. The specific observations analyzed can be accessed via \dataset[doi:10.17909/4012-2q76]{[https://doi.org/10.17909/4012-2q76]}.
\end{acknowledgments}





%
\facilities{\textit{HST}(STIS)}

\software{ 
            Grizli \citep{2019ascl.soft05001B},
            EMCEE \citep{emcee},
            GALFIT \citep{2002AJ....124..266P},
            VorBin \citep{2003MNRAS.342..345C} 
            }

\clearpage






\bibliography{Main}

@ARTICLE{1996A&AS..117..393B,
       author = {{Bertin}, E. and {Arnouts}, S.},
        title = "{SExtractor: Software for source extraction.}",
      journal = {\aaps},
         year = "1996",
        month = "Jun",
       volume = {117},
        pages = {393-404},
          doi = {10.1051/aas:1996164},
       adsurl = {https://ui.adsabs.harvard.edu/abs/1996A&AS..117..393B}
}

@misc{Grizli,
       author = {{Brammer}, Gabe and {Matharu}, Jasleen},
        title = "{gbrammer/grizli: Release 2021}",
         year = 2021,
        month = jun,
          eid = {10.5281/zenodo.5012699},
          doi = {10.5281/zenodo.5012699},
      version = {1.3.2},
    publisher = {Zenodo},
       adsurl = {https://ui.adsabs.harvard.edu/abs/2021zndo...5012699B}
}

@article{merlinTPHOTVersion202016,
	title = {T-{PHOT} version 2.0: {Improved} algorithms for background subtraction, local convolution, kernel registration, and new options},
	volume = {595},
	issn = {0004-6361},
	shorttitle = {T-{PHOT} version 2.0},
	url = {https://ui.adsabs.harvard.edu/abs/2016A&A...595A..97M},
	doi = {10.1051/0004-6361/201628751},
	urldate = {2024-02-23},
	journal = {Astronomy and Astrophysics},
	author = {Merlin, E. and Bourne, N. and Castellano, M. and Ferguson, H. C. and Wang, T. and Derriere, S. and Dunlop, J. S. and Elbaz, D. and Fontana, A.},
	month = nov,
	year = {2016},
	note = {ADS Bibcode: 2016A\&A...595A..97M},
	pages = {A97},
}

@ARTICLE{2011ApJ...736..104J,
       author = {{Juneau}, St{\'e}phanie and {Dickinson}, Mark and {Alexander}, David M. and {Salim}, Samir},
        title = "{A New Diagnostic of Active Galactic Nuclei: Revealing Highly Absorbed Systems at Redshift >0.3}",
      journal = {\apj},
         year = 2011,
        month = aug,
       volume = {736},
       number = {2},
          eid = {104},
        pages = {104},
          doi = {10.1088/0004-637X/736/2/104},
archivePrefix = {arXiv},
       eprint = {1105.3194},
 primaryClass = {astro-ph.CO},
       adsurl = {https://ui.adsabs.harvard.edu/abs/2011ApJ...736..104J}
}

@ARTICLE{2014ApJ...788...88J,
       author = {{Juneau}, St{\'e}phanie and {Bournaud}, Fr{\'e}d{\'e}ric and {Charlot}, St{\'e}phane and {Daddi}, Emanuele and {Elbaz}, David and {Trump}, Jonathan R. and {Brinchmann}, Jarle and {Dickinson}, Mark and {Duc}, Pierre-Alain and {Gobat}, Raphael and {Jean-Baptiste}, Ingrid and {Le Floc'h}, {\'E}meric and {Lehnert}, M.~D. and {Pacifici}, Camilla and {Pannella}, Maurilio and {Schreiber}, Corentin},
        title = "{Active Galactic Nuclei Emission Line Diagnostics and the Mass-Metallicity Relation up to Redshift z \raisebox{-0.5ex}\textasciitilde 2: The Impact of Selection Effects and Evolution}",
      journal = {\apj},
         year = 2014,
        month = jun,
       volume = {788},
       number = {1},
          eid = {88},
        pages = {88},
          doi = {10.1088/0004-637X/788/1/88},
archivePrefix = {arXiv},
       eprint = {1403.6832},
 primaryClass = {astro-ph.GA},
       adsurl = {https://ui.adsabs.harvard.edu/abs/2014ApJ...788...88J}
}

@ARTICLE{2015ApJ...801...35C,
       author = {{Coil}, Alison L. and {Aird}, James and {Reddy}, Naveen and {Shapley}, Alice E. and {Kriek}, Mariska and {Siana}, Brian and {Mobasher}, Bahram and {Freeman}, William R. and {Price}, Sedona H. and {Shivaei}, Irene},
        title = "{The MOSDEF Survey: Optical Active Galactic Nucleus Diagnostics at z \raisebox{-0.5ex}\textasciitilde 2.3}",
      journal = {\apj},
         year = 2015,
        month = mar,
       volume = {801},
       number = {1},
          eid = {35},
        pages = {35},
          doi = {10.1088/0004-637X/801/1/35},
archivePrefix = {arXiv},
       eprint = {1409.6522},
 primaryClass = {astro-ph.GA},
       adsurl = {https://ui.adsabs.harvard.edu/abs/2015ApJ...801...35C}
}

@ARTICLE{2018MNRAS.480.4379C,
       author = {{Carnall}, A.~C. and {McLure}, R.~J. and {Dunlop}, J.~S. and {Dav{\'e}}, R.},
        title = "{Inferring the star formation histories of massive quiescent galaxies with BAGPIPES: evidence for multiple quenching mechanisms}",
      journal = {\mnras},
         year = 2018,
        month = nov,
       volume = {480},
       number = {4},
        pages = {4379-4401},
          doi = {10.1093/mnras/sty2169},
archivePrefix = {arXiv},
       eprint = {1712.04452},
 primaryClass = {astro-ph.GA},
       adsurl = {https://ui.adsabs.harvard.edu/abs/2018MNRAS.480.4379C}
}

@ARTICLE{2003PASP..115..763C,
       author = {{Chabrier}, Gilles},
        title = "{Galactic Stellar and Substellar Initial Mass Function}",
      journal = {\pasp},
         year = 2003,
        month = jul,
       volume = {115},
       number = {809},
        pages = {763-795},
          doi = {10.1086/376392},
archivePrefix = {arXiv},
       eprint = {astro-ph/0304382},
 primaryClass = {astro-ph},
       adsurl = {https://ui.adsabs.harvard.edu/abs/2003PASP..115..763C}
}

@ARTICLE{2003MNRAS.344.1000B,
       author = {{Bruzual}, G. and {Charlot}, S.},
        title = "{Stellar population synthesis at the resolution of 2003}",
      journal = {\mnras},
         year = 2003,
        month = oct,
       volume = {344},
       number = {4},
        pages = {1000-1028},
          doi = {10.1046/j.1365-8711.2003.06897.x},
archivePrefix = {arXiv},
       eprint = {astro-ph/0309134},
 primaryClass = {astro-ph},
       adsurl = {https://ui.adsabs.harvard.edu/abs/2003MNRAS.344.1000B}
}

@ARTICLE{Av2000C,
       author = {{Calzetti}, Daniela and {Armus}, Lee and {Bohlin}, Ralph C. and {Kinney}, Anne L. and {Koornneef}, Jan and {Storchi-Bergmann}, Thaisa},
        title = "{The Dust Content and Opacity of Actively Star-forming Galaxies}",
      journal = {\apj},
         year = 2000,
        month = apr,
       volume = {533},
       number = {2},
        pages = {682-695},
          doi = {10.1086/308692},
archivePrefix = {arXiv},
       eprint = {astro-ph/9911459},
 primaryClass = {astro-ph},
       adsurl = {https://ui.adsabs.harvard.edu/abs/2000ApJ...533..682C}
}

@ARTICLE{wang2017,
       author = {{Wang}, Xin and {Jones}, Tucker A. and {Treu}, Tommaso and {Morishita}, Takahiro and {Abramson}, Louis E. and {Brammer}, Gabriel B. and {Huang}, Kuang-Han and {Malkan}, Matthew A. and {Schmidt}, Kasper B. and {Fontana}, Adriano and {Grillo}, Claudio and {Henry}, Alaina L. and {Karman}, Wouter and {Kelly}, Patrick L. and {Mason}, Charlotte A. and {Mercurio}, Amata and {Rosati}, Piero and {Sharon}, Keren and {Trenti}, Michele and {Vulcani}, Benedetta},
        title = "{The Grism Lens-amplified Survey from Space (GLASS). X. Sub-kiloparsec Resolution Gas-phase Metallicity Maps at Cosmic Noon behind the Hubble Frontier Fields Cluster MACS1149.6+2223}",
      journal = {\apj},
         year = 2017,
        month = mar,
       volume = {837},
       number = {1},
          eid = {89},
        pages = {89},
          doi = {10.3847/1538-4357/aa603c},
archivePrefix = {arXiv},
       eprint = {1610.07558},
 primaryClass = {astro-ph.GA},
       adsurl = {https://ui.adsabs.harvard.edu/abs/2017ApJ...837...89W}
}

@ARTICLE{wang2019,
       author = {{Wang}, Xin and {Jones}, Tucker A. and {Treu}, Tommaso and {Hirtenstein}, Jessie and {Brammer}, Gabriel B. and {Daddi}, Emanuele and {Meng}, Xiao-Lei and {Morishita}, Takahiro and {Abramson}, Louis E. and {Henry}, Alaina L. and {Peng}, Ying-jie and {Schmidt}, Kasper B. and {Sharon}, Keren and {Trenti}, Michele and {Vulcani}, Benedetta},
        title = "{Discovery of Strongly Inverted Metallicity Gradients in Dwarf Galaxies at z {\ensuremath{\sim}} 2}",
      journal = {\apj},
         year = 2019,
        month = sep,
       volume = {882},
       number = {2},
          eid = {94},
        pages = {94},
          doi = {10.3847/1538-4357/ab3861},
archivePrefix = {arXiv},
       eprint = {1808.08800},
 primaryClass = {astro-ph.GA},
       adsurl = {https://ui.adsabs.harvard.edu/abs/2019ApJ...882...94W}
}

@ARTICLE{wang2020b,
       author = {{Wang}, Xin and {Jones}, Tucker A. and {Treu}, Tommaso and {Daddi}, Emanuele and {Brammer}, Gabriel B. and {Sharon}, Keren and {Morishita}, Takahiro and {Abramson}, Louis E. and {Colbert}, James W. and {Henry}, Alaina L. and {Hopkins}, Philip F. and {Malkan}, Matthew A. and {Schmidt}, Kasper B. and {Teplitz}, Harry I. and {Vulcani}, Benedetta},
        title = "{A Census of Sub-kiloparsec Resolution Metallicity Gradients in Star-forming Galaxies at Cosmic Noon from HST Slitless Spectroscopy}",
      journal = {\apj},
         year = 2020,
        month = sep,
       volume = {900},
       number = {2},
          eid = {183},
        pages = {183},
          doi = {10.3847/1538-4357/abacce},
archivePrefix = {arXiv},
       eprint = {1911.09841},
 primaryClass = {astro-ph.GA},
       adsurl = {https://ui.adsabs.harvard.edu/abs/2020ApJ...900..183W}
}

@ARTICLE{Wang_2022,
       author = {{Wang}, Xin and {Li}, Zihao and {Cai}, Zheng and {Shi}, Dong Dong and {Fan}, Xiaohui and {Zheng}, Xian Zhong and {Bian}, Fuyan and {Teplitz}, Harry I. and {Alavi}, Anahita and {Colbert}, James and {Henry}, Alaina L. and {Malkan}, Matthew A.},
        title = "{The Mass-Metallicity Relation at Cosmic Noon in Overdense Environments: First Results from the MAMMOTH-Grism HST Slitless Spectroscopic Survey}",
      journal = {\apj},
         year = 2022,
        month = feb,
       volume = {926},
       number = {1},
          eid = {70},
        pages = {70},
          doi = {10.3847/1538-4357/ac3974},
archivePrefix = {arXiv},
       eprint = {2108.06373},
 primaryClass = {astro-ph.GA},
       adsurl = {https://ui.adsabs.harvard.edu/abs/2022ApJ...926...70W}
}

@ARTICLE{2003MNRAS.342..345C,
       author = {{Cappellari}, Michele and {Copin}, Yannick},
        title = "{Adaptive spatial binning of integral-field spectroscopic data using Voronoi tessellations}",
      journal = {\mnras},
         year = 2003,
        month = jun,
       volume = {342},
       number = {2},
        pages = {345-354},
          doi = {10.1046/j.1365-8711.2003.06541.x},
archivePrefix = {arXiv},
       eprint = {astro-ph/0302262},
 primaryClass = {astro-ph},
       adsurl = {https://ui.adsabs.harvard.edu/abs/2003MNRAS.342..345C}
}

@ARTICLE{2018ApJ...859..175B,
       author = {{Bian}, Fuyan and {Kewley}, Lisa J. and {Dopita}, Michael A.},
        title = "{{\textquotedblleft}Direct{\textquotedblright} Gas-phase Metallicity in Local Analogs of High-redshift Galaxies: Empirical Metallicity Calibrations for High-redshift Star-forming Galaxies}",
      journal = {\apj},
         year = 2018,
        month = jun,
       volume = {859},
       number = {2},
          eid = {175},
        pages = {175},
          doi = {10.3847/1538-4357/aabd74},
archivePrefix = {arXiv},
       eprint = {1805.08224},
 primaryClass = {astro-ph.GA},
       adsurl = {https://ui.adsabs.harvard.edu/abs/2018ApJ...859..175B}
}

@software{emcee,
       author = {{Foreman-Mackey}, Daniel and {Conley}, Alex and {Meierjurgen Farr}, Will and {Hogg}, David W. and {Lang}, Dustin and {Marshall}, Phil and {Price-Whelan}, Adrian and {Sanders}, Jeremy and {Zuntz}, Joe},
        title = "{emcee: The MCMC Hammer}",
 howpublished = {Astrophysics Source Code Library, record ascl:1303.002},
         year = 2013,
        month = mar,
          eid = {ascl:1303.002},
       adsurl = {https://ui.adsabs.harvard.edu/abs/2013ascl.soft03002F}
}

@ARTICLE{2002AJ....124..266P,
       author = {{Peng}, Chien Y. and {Ho}, Luis C. and {Impey}, Chris D. and {Rix}, Hans-Walter},
        title = "{Detailed Structural Decomposition of Galaxy Images}",
      journal = {\aj},
         year = 2002,
        month = jul,
       volume = {124},
       number = {1},
        pages = {266-293},
          doi = {10.1086/340952},
archivePrefix = {arXiv},
       eprint = {astro-ph/0204182},
 primaryClass = {astro-ph},
       adsurl = {https://ui.adsabs.harvard.edu/abs/2002AJ....124..266P}
}

@ARTICLE{2016ApJ...820...84L,
       author = {{Leethochawalit}, Nicha and {Jones}, Tucker A. and {Ellis}, Richard S. and {Stark}, Daniel P. and {Richard}, Johan and {Zitrin}, Adi and {Auger}, Matthew},
        title = "{A Keck Adaptive Optics Survey of a Representative Sample of Gravitationally Lensed Star-forming Galaxies: High Spatial Resolution Studies of Kinematics and Metallicity Gradients}",
      journal = {\apj},
         year = 2016,
        month = apr,
       volume = {820},
       number = {2},
          eid = {84},
        pages = {84},
          doi = {10.3847/0004-637X/820/2/84},
archivePrefix = {arXiv},
       eprint = {1509.01279},
 primaryClass = {astro-ph.GA},
       adsurl = {https://ui.adsabs.harvard.edu/abs/2016ApJ...820...84L}
}

@ARTICLE{2016ApJ...827...74W,
       author = {{Wuyts}, Eva and {Wisnioski}, Emily and {Fossati}, Matteo and {F{\"o}rster Schreiber}, Natascha M. and {Genzel}, Reinhard and {Davies}, Ric and {Mendel}, J. Trevor and {Naab}, Thorsten and {R{\"o}ttgers}, Bernhard and {Wilman}, David J. and {Wuyts}, Stijn and {Bandara}, Kaushala and {Beifiori}, Alessandra and {Belli}, Sirio and {Bender}, Ralf and {Brammer}, Gabriel B. and {Burkert}, Andreas and {Chan}, Jeffrey and {Galametz}, Audrey and {Kulkarni}, Sandesh K. and {Lang}, Philipp and {Lutz}, Dieter and {Momcheva}, Ivelina G. and {Nelson}, Erica J. and {Rosario}, David and {Saglia}, Roberto P. and {Seitz}, Stella and {Tacconi}, Linda J. and {Tadaki}, Ken-ichi and {{\"U}bler}, Hannah and {van Dokkum}, Pieter},
        title = "{The Evolution of Metallicity and Metallicity Gradients from z = 2.7 to 0.6 with KMOS$^{3D}$}",
      journal = {\apj},
         year = 2016,
        month = aug,
       volume = {827},
       number = {1},
          eid = {74},
        pages = {74},
          doi = {10.3847/0004-637X/827/1/74},
archivePrefix = {arXiv},
       eprint = {1603.01139},
 primaryClass = {astro-ph.GA},
       adsurl = {https://ui.adsabs.harvard.edu/abs/2016ApJ...827...74W}
}

@ARTICLE{2020MNRAS.492..821C,
       author = {{Curti}, Mirko and {Maiolino}, Roberto and {Cirasuolo}, Michele and {Mannucci}, Filippo and {Williams}, Rebecca J. and {Auger}, Matt and {Mercurio}, Amata and {Hayden-Pawson}, Connor and {Cresci}, Giovanni and {Marconi}, Alessandro and {Belfiore}, Francesco and {Cappellari}, Michele and {Cicone}, Claudia and {Cullen}, Fergus and {Meneghetti}, Massimo and {Ota}, Kazuaki and {Peng}, Yingjie and {Pettini}, Max and {Swinbank}, Mark and {Troncoso}, Paulina},
        title = "{The KLEVER Survey: spatially resolved metallicity maps and gradients in a sample of 1.2 < z < 2.5 lensed galaxies}",
      journal = {\mnras},
         year = 2020,
        month = feb,
       volume = {492},
       number = {1},
        pages = {821-842},
          doi = {10.1093/mnras/stz3379},
archivePrefix = {arXiv},
       eprint = {1910.13451},
 primaryClass = {astro-ph.GA},
       adsurl = {https://ui.adsabs.harvard.edu/abs/2020MNRAS.492..821C}
}

@ARTICLE{2014MNRAS.443.3643P,
       author = {{Peng}, Ying-jie and {Maiolino}, Roberto},
        title = "{From haloes to Galaxies - I. The dynamics of the gas regulator model and the implied cosmic sSFR history}",
      journal = {\mnras},
         year = 2014,
        month = oct,
       volume = {443},
       number = {4},
        pages = {3643-3664},
          doi = {10.1093/mnras/stu1288},
archivePrefix = {arXiv},
       eprint = {1402.5964},
 primaryClass = {astro-ph.CO},
       adsurl = {https://ui.adsabs.harvard.edu/abs/2014MNRAS.443.3643P}
}

@article{cresciGasAccretionOrigin2010,
  title = {Gas Accretion as the Origin of Chemical Abundance Gradients in Distant Galaxies},
  author = {Cresci, G. and Mannucci, F. and Maiolino, R. and Marconi, A. and Gnerucci, A. and Magrini, L.},
  year = 2010,
  month = oct,
  journal = {Nature},
  volume = {467},
  number = {7317},
  pages = {811--813},
  publisher = {Nature Publishing Group},
  issn = {1476-4687},
  doi = {10.1038/nature09451},
  urldate = {2025-12-24},
  copyright = {2010 Springer Nature Limited},
  langid = {english},
}

@article{hoMetallicityGradientsLocal2015,
  title = {Metallicity Gradients in Local Field Star-Forming Galaxies: Insights on Inflows, Outflows, and the Coevolution of Gas, Stars and Metals},
  shorttitle = {Metallicity Gradients in Local Field Star-Forming Galaxies},
  author = {Ho, I. -Ting and Kudritzki, Rolf-Peter and Kewley, Lisa J. and Zahid, H. Jabran and Dopita, Michael A. and Bresolin, Fabio and Rupke, David S. N.},
  year = 2015,
  month = apr,
  journal = {Monthly Notices of the Royal Astronomical Society},
  volume = {448},
  pages = {2030--2054},
  publisher = {OUP},
  issn = {0035-8711},
  doi = {10.1093/mnras/stv067},
  urldate = {2025-01-13},
}

@article{sanchezCharacteristicOxygenAbundance2014,
  title = {A Characteristic Oxygen Abundance Gradient in Galaxy Disks Unveiled with {{CALIFA}}},
  author = {S{\'a}nchez, S. F. and {Rosales-Ortega}, F. F. and {Iglesias-P{\'a}ramo}, J. and Moll{\'a}, M. and {Barrera-Ballesteros}, J. and Marino, R. A. and P{\'e}rez, E. and {S{\'a}nchez-Blazquez}, P. and Gonz{\'a}lez Delgado, R. and Cid Fernandes, R. and {De Lorenzo-C{\'a}ceres}, A. and {Mendez-Abreu}, J. and Galbany, L. and {Falcon-Barroso}, J. and {Miralles-Caballero}, D. and Husemann, B. and {Garc{\'i}a-Benito}, R. and Mast, D. and Walcher, C. J. and Gil De Paz, A. and {Garc{\'i}a-Lorenzo}, B. and Jungwiert, B. and V{\'i}lchez, J. M. and J{\'i}lkov{\'a}, Lucie and Lyubenova, M. and {Cortijo-Ferrero}, C. and D{\'i}az, A. I. and Wisotzki, L. and M{\'a}rquez, I. and {Bland-Hawthorn}, J. and Ellis, S. and Van De Ven, G. and Jahnke, K. and Papaderos, P. and Gomes, J. M. and Mendoza, M. A. and {L{\'o}pez-S{\'a}nchez}, {\'A}. R. and {The CALIFA collaboration}},
  year = 2014,
  month = mar,
  journal = {Astronomy \& Astrophysics},
  volume = {563},
  pages = {A49},
  issn = {0004-6361, 1432-0746},
  doi = {10.1051/0004-6361/201322343},
  urldate = {2025-12-24},
  langid = {english},
}

@article{stanghelliniGALACTICSTRUCTURECHEMICAL2010,
  title = {{{THE GALACTIC STRUCTURE AND CHEMICAL EVOLUTION TRACED BY THE POPULATION OF PLANETARY NEBULAE}}},
  author = {Stanghellini, Letizia and Haywood, Misha},
  year = 2010,
  month = apr,
  journal = {The Astrophysical Journal},
  volume = {714},
  number = {2},
  pages = {1096},
  publisher = {The American Astronomical Society},
  issn = {0004-637X},
  doi = {10.1088/0004-637X/714/2/1096},
  urldate = {2025-12-24},
  langid = {english}
}

@article{luckDISTRIBUTIONELEMENTSGALACTIC2011,
  title = {{{THE DISTRIBUTION OF THE ELEMENTS IN THE GALACTIC DISK}}. {{II}}. {{AZIMUTHAL AND RADIAL VARIATION IN ABUNDANCES FROM CEPHEIDS}}},
  author = {Luck, R. E. and Andrievsky, S. M. and Kovtyukh, V. V. and Gieren, W. and Graczyk, D.},
  year = 2011,
  month = jul,
  journal = {The Astronomical Journal},
  volume = {142},
  number = {2},
  pages = {51},
  publisher = {The American Astronomical Society},
  issn = {1538-3881},
  doi = {10.1088/0004-6256/142/2/51},
  urldate = {2025-12-24},
  langid = {english},
}

@article{belfioreSDSSIVMaNGA2017,
  title = {{{SDSS IV MaNGA}} -- Metallicity and Nitrogen Abundance Gradients in Local Galaxies},
  author = {Belfiore, Francesco and Maiolino, Roberto and Tremonti, Christy and S{\'a}nchez, Sebastian F. and Bundy, Kevin and Bershady, Matthew and Westfall, Kyle and Lin, Lihwai and Drory, Niv and Boquien, M{\'e}d{\'e}ric and Thomas, Daniel and Brinkmann, Jonathan},
  year = 2017,
  month = jul,
  journal = {Monthly Notices of the Royal Astronomical Society},
  volume = {469},
  number = {1},
  pages = {151--170},
  issn = {0035-8711},
  doi = {10.1093/mnras/stx789},
  urldate = {2025-12-24},
}

@article{maiolinoReMetallicaCosmic2019,
  title = {De Re Metallica: The Cosmic Chemical Evolution of Galaxies},
  shorttitle = {De Re Metallica},
  author = {Maiolino, R. and Mannucci, F.},
  year = 2019,
  month = feb,
  journal = {The Astronomy and Astrophysics Review},
  volume = {27},
  number = {1},
  pages = {3},
  issn = {1432-0754},
  doi = {10.1007/s00159-018-0112-2},
  urldate = {2025-12-24},
  langid = {english},
}

@article{liFirstCensusGasphase2022,
  title = {First {{Census}} of {{Gas-phase Metallicity Gradients}} of {{Star-forming Galaxies}} in {{Overdense Environments}} at {{Cosmic Noon}}},
  author = {Li, Zihao and Wang, Xin and Cai, Zheng and Shi, Dong Dong and Fan, Xiaohui and Zheng, Xian Zhong and Malkan, Matthew A. and Teplitz, Harry I. and Henry, Alaina L. and Bian, Fuyan and Colbert, James},
  year = 2022,
  month = apr,
  journal = {The Astrophysical Journal Letters},
  volume = {929},
  number = {1},
  pages = {L8},
  publisher = {The American Astronomical Society},
  issn = {2041-8205},
  doi = {10.3847/2041-8213/ac626f},
  urldate = {2025-12-24},
  langid = {english},
}

@article{venturiGasphaseMetallicityGradients2024,
  title = {Gas-Phase Metallicity Gradients in Galaxies at \$z \textbackslash sim 6-8\$},
  author = {Venturi, G. and Carniani, S. and Parlanti, E. and Kohandel, M. and Curti, M. and Pallottini, A. and Vallini, L. and Arribas, S. and Bunker, A. J. and Cameron, A. J. and Castellano, M. and Ferrara, A. and Fontana, A. and Gallerani, S. and Gelli, V. and Maiolino, R. and Ntormousi, E. and Pacifici, C. and Pentericci, L. and Salvadori, S. and Vanzella, E.},
  year = 2024,
  month = nov,
  journal = {Astronomy \& Astrophysics},
  volume = {691},
  eprint = {2403.03977},
  primaryclass = {astro-ph},
  pages = {A19},
  issn = {0004-6361, 1432-0746},
  doi = {10.1051/0004-6361/202449855},
  urldate = {2025-04-08},
  archiveprefix = {arXiv},
}

@article{li13BillionYear2025,
  title = {A 13 {{Billion Year View}} of {{Galaxy Growth}}: {{Metallicity Gradient Evolution}} from the {{Local Universe}} to z = 9 with {{JWST}} and {{Archival Surveys}}},
  shorttitle = {A 13 {{Billion Year View}} of {{Galaxy Growth}}},
  author = {Li, Zihao and Cai, Zheng and Wang, Xin and Li, Zhaozhou and Dekel, Avishai and Sarkar, Kartick C. and Ba{\~n}ados, Eduardo and Bian, Fuyan and Bhowmick, Aklant K. and Blecha, Laura and Bosman, Sarah E. I. and Champagne, Jaclyn B. and Fan, Xiaohui and {Golden-Marx}, Emmet and Jun, Hyunsung D. and Li, Mingyu and Lin, Xiaojing and Liu, Weizhe and Sun, Fengwu and Trebitsch, Maxime and Walter, Fabian and Wang, Feige and Wu, Yunjing and Yang, Jinyi and Zhang, Huanian and Zhang, Shiwu and Zhuang, Mingyang and Zou, Siwei},
  year = 2025,
  month = sep,
  journal = {The Astrophysical Journal Supplement Series},
  volume = {280},
  number = {2},
  pages = {62},
  publisher = {The American Astronomical Society},
  issn = {0067-0049},
  doi = {10.3847/1538-4365/adfa70},
  urldate = {2025-12-29},
  langid = {english},
}

@article{juMSA3DMetallicityGradients2025,
  title = {{{MSA-3D}}: {{Metallicity Gradients}} in {{Galaxies}} at z {$\sim$} 1 with {{JWST}}/{{NIRSpec Slit-stepping Spectroscopy}}},
  shorttitle = {{{MSA-3D}}},
  author = {Ju, Mengting and Wang, Xin and Jones, Tucker and Bari{\v s}i{\'c}, Ivana and Nanayakkara, Themiya and Bundy, Kevin and {Faucher-Gigu{\`e}re}, Claude-Andr{\'e} and Feng, Shuai and Glazebrook, Karl and Henry, Alaina and Malkan, Matthew A. and Obreschkow, Danail and Roy, Namrata and Sanders, Ryan L. and Sun, Xunda and Treu, Tommaso and Zhou, Qianqiao},
  year = 2025,
  month = jan,
  journal = {The Astrophysical Journal Letters},
  volume = {978},
  number = {2},
  pages = {L39},
  publisher = {The American Astronomical Society},
  issn = {2041-8205},
  doi = {10.3847/2041-8213/ada150},
  urldate = {2025-12-29},
  langid = {english},
}

@article{sunPhysicalOriginPositive2025,
  title = {The {{Physical Origin}} of {{Positive Metallicity Radial Gradients}} in {{High-redshift Galaxies}}: {{Insights}} from the {{FIRE-2 Cosmological Hydrodynamic Simulations}}},
  shorttitle = {The {{Physical Origin}} of {{Positive Metallicity Radial Gradients}} in {{High-redshift Galaxies}}},
  author = {Sun, Xunda and Wang, Xin and Ma, Xiangcheng and Wang, Kai and Wetzel, Andrew and {Faucher-Gigu{\`e}re}, Claude-Andr{\'e} and Hopkins, Philip F. and Kere{\v s}, Du{\v s}an and Graf, Russell L. and Marszewski, Andrew and Stern, Jonathan and Sun, Guochao and Sun, Lei and Thyme, Keyer},
  year = 2025,
  month = jun,
  journal = {The Astrophysical Journal},
  volume = {986},
  number = {2},
  pages = {179},
  publisher = {The American Astronomical Society},
  issn = {0004-637X},
  doi = {10.3847/1538-4357/addab5},
  urldate = {2025-12-29},
  langid = {english},
}

@article{gibsonConstrainingSubgridPhysics2013,
  title = {Constraining Sub-Grid Physics with High-Redshift Spatially-Resolved Metallicity Distributions},
  author = {Gibson, B. K. and Pilkington, K. and Brook, C. B. and Stinson, G. S. and Bailin, J.},
  year = 2013,
  month = jun,
  journal = {Astronomy \& Astrophysics},
  volume = {554},
  pages = {A47},
  publisher = {EDP Sciences},
  issn = {0004-6361, 1432-0746},
  doi = {10.1051/0004-6361/201321239},
  urldate = {2025-12-29},
  copyright = {\copyright{} ESO, 2013},
  langid = {english},
}

@article{queyrelMASSIVMassAssembly2012,
  title = {{{MASSIV}}: {{Mass Assembly Survey}} with {{SINFONI}} in {{VVDS}} - {{III}}. {{Evidence}} for Positive Metallicity Gradients in z \textasciitilde{} 1.2 Star-Forming Galaxies},
  shorttitle = {{{MASSIV}}},
  author = {Queyrel, J. and Contini, T. and {Kissler-Patig}, M. and Epinat, B. and Amram, P. and Garilli, B. and F{\`e}vre, O. Le and Moultaka, J. and Paioro, L. and Tasca, L. and Tresse, L. and Vergani, D. and {L{\'o}pez-Sanjuan}, C. and {Perez-Montero}, E.},
  year = 2012,
  month = mar,
  journal = {Astronomy \& Astrophysics},
  volume = {539},
  pages = {A93},
  publisher = {EDP Sciences},
  issn = {0004-6361, 1432-0746},
  doi = {10.1051/0004-6361/201117718},
  urldate = {2025-12-29},
  copyright = {\copyright{} ESO, 2012},
  langid = {english},
}

@article{hemlerGasphaseMetallicityGradients2021,
  title = {Gas-Phase Metallicity Gradients of {{TNG50}} Star-Forming Galaxies},
  author = {Hemler, Z S and Torrey, Paul and Qi, Jia and Hernquist, Lars and Vogelsberger, Mark and Ma, Xiangcheng and Kewley, Lisa J and Nelson, Dylan and Pillepich, Annalisa and Pakmor, R{\"u}diger and Marinacci, Federico},
  year = 2021,
  month = sep,
  journal = {Monthly Notices of the Royal Astronomical Society},
  volume = {506},
  number = {2},
  pages = {3024--3048},
  issn = {0035-8711},
  doi = {10.1093/mnras/stab1803},
  urldate = {2025-12-29}
}

@article{tisseraEvolutionOxygenAbundance2022,
  title = {The Evolution of the Oxygen Abundance Gradients in Star-Forming Galaxies in the Eagle Simulations},
  author = {Tissera, Patricia B and {Rosas-Guevara}, Yetli and Sillero, Emanuel and Pedrosa, Susana E and Theuns, Tom and Bignone, Lucas},
  year = 2022,
  month = apr,
  journal = {Monthly Notices of the Royal Astronomical Society},
  volume = {511},
  number = {2},
  pages = {1667--1684},
  issn = {0035-8711},
  doi = {10.1093/mnras/stab3644},
  urldate = {2025-12-29}
}

@article{jonesORIGINEVOLUTIONMETALLICITY2013,
  title = {{{THE ORIGIN AND EVOLUTION OF METALLICITY GRADIENTS}}: {{PROBING THE MODE OF MASS ASSEMBLY AT}} z {$\simeq$} 2},
  shorttitle = {{{THE ORIGIN AND EVOLUTION OF METALLICITY GRADIENTS}}},
  author = {Jones, Tucker and Ellis, Richard S. and Richard, Johan and Jullo, Eric},
  year = 2013,
  month = feb,
  journal = {The Astrophysical Journal},
  volume = {765},
  number = {1},
  pages = {48},
  publisher = {The American Astronomical Society},
  issn = {0004-637X},
  doi = {10.1088/0004-637X/765/1/48},
  urldate = {2025-12-29},
  langid = {english},
}

@article{swinbankPropertiesStarformingInterstellar2012,
  title = {The Properties of the Star-Forming Interstellar Medium at z = 0.84--2.23 from {{HiZELS}}: Mapping the Internal Dynamics and Metallicity Gradients in High-Redshift Disc Galaxies},
  shorttitle = {The Properties of the Star-Forming Interstellar Medium at z = 0.84--2.23 from {{HiZELS}}},
  author = {Swinbank, A. M. and Sobral, D. and Smail, Ian and Geach, J. E. and Best, P. N. and McCarthy, I. G. and Crain, R. A. and Theuns, T.},
  year = 2012,
  month = oct,
  journal = {Monthly Notices of the Royal Astronomical Society},
  volume = {426},
  number = {2},
  pages = {935--950},
  issn = {0035-8711},
  doi = {10.1111/j.1365-2966.2012.21774.x},
  urldate = {2025-12-29}
}

@article{molinaSINFONIHiZELSDynamicsMerger2017,
  title = {{{SINFONI-HiZELS}}: The Dynamics, Merger Rates and Metallicity Gradients of `Typical' Star-Forming Galaxies at z~=~0.8--2.2},
  shorttitle = {{{SINFONI-HiZELS}}},
  author = {Molina, J. and Ibar, Edo and Swinbank, A. M. and Sobral, D. and Best, P. N. and Smail, I. and Escala, A. and Cirasuolo, M.},
  year = 2017,
  month = apr,
  journal = {Monthly Notices of the Royal Astronomical Society},
  volume = {466},
  number = {1},
  pages = {892--905},
  issn = {0035-8711},
  doi = {10.1093/mnras/stw3120},
  urldate = {2025-12-29}
}

@article{cartonFirstGasphaseMetallicity2018,
  title = {First Gas-Phase Metallicity Gradients of 0.1 {$\lessequivlnt$} z {$\lessequivlnt$} 0.8 Galaxies with {{MUSE}}},
  author = {Carton, David and Brinchmann, Jarle and Contini, Thierry and Epinat, Beno{\^i}t and Finley, Hayley and Richard, Johan and Patr{\'i}cio, Vera and Schaye, Joop and Nanayakkara, Themiya and Weilbacher, Peter M and Wisotzki, Lutz},
  year = 2018,
  month = aug,
  journal = {Monthly Notices of the Royal Astronomical Society},
  volume = {478},
  number = {4},
  pages = {4293--4316},
  issn = {0035-8711},
  doi = {10.1093/mnras/sty1343},
  urldate = {2025-12-29}
}

@article{schreiberSINSZCSINFSurvey2018,
  title = {The {{SINS}}/{{zC-SINF Survey}} of z {$\sim$} 2 {{Galaxy Kinematics}}: {{SINFONI Adaptive Optics}}--Assisted {{Data}} and {{Kiloparsec-scale Emission-line Properties}}{$\ast$}},
  shorttitle = {The {{SINS}}/{{zC-SINF Survey}} of z {$\sim$} 2 {{Galaxy Kinematics}}},
  author = {Schreiber, N. M. F{\"o}rster and Renzini, A. and Mancini, C. and Genzel, R. and Bouch{\'e}, N. and Cresci, G. and Hicks, E. K. S. and Lilly, S. J. and Peng, Y. and Burkert, A. and Carollo, C. M. and Cimatti, A. and Daddi, E. and Davies, R. I. and Genel, S. and Kurk, J. D. and Lang, P. and Lutz, D. and Mainieri, V. and McCracken, H. J. and Mignoli, M. and Naab, T. and Oesch, P. and Pozzetti, L. and Scodeggio, M. and Griffin, K. Shapiro and Shapley, A. E. and Sternberg, A. and Tacchella, S. and Tacconi, L. J. and Wuyts, S. and Zamorani, G.},
  year = 2018,
  month = oct,
  journal = {The Astrophysical Journal Supplement Series},
  volume = {238},
  number = {2},
  pages = {21},
  publisher = {The American Astronomical Society},
  issn = {0067-0049},
  doi = {10.3847/1538-4365/aadd49},
  urldate = {2025-12-29},
  langid = {english}
}

@article{patricioResolvedScalingRelations2019,
  title = {Resolved Scaling Relations and Metallicity Gradients on Sub-Kiloparsec Scales at z {$\approx$} 1},
  author = {Patr{\'i}cio, V and Richard, J and Carton, D and P{\'e}roux, C and Contini, T and Brinchmann, J and Schaye, J and Weilbacher, P M and Nanayakkara, T and Maseda, M and Mahler, G and Wisotzki, L},
  year = 2019,
  month = oct,
  journal = {Monthly Notices of the Royal Astronomical Society},
  volume = {489},
  number = {1},
  pages = {224--240},
  issn = {0035-8711},
  doi = {10.1093/mnras/stz2114},
  urldate = {2025-12-29}
}

@article{simonsCLEARGasphaseMetallicity2021,
  title = {{{CLEAR}}: {{The Gas-phase Metallicity Gradients}} of {{Star-forming Galaxies}} at 0.6 {$<$} z {$<$} 2.6},
  shorttitle = {{{CLEAR}}},
  author = {Simons, Raymond C. and Papovich, Casey and Momcheva, Ivelina and Trump, Jonathan R. and Brammer, Gabriel and {Estrada-Carpenter}, Vicente and Backhaus, Bren E. and Cleri, Nikko J. and Finkelstein, Steven L. and Giavalisco, Mauro and Ji, Zhiyuan and Jung, Intae and Matharu, Jasleen and Weiner, Benjamin},
  year = 2021,
  month = dec,
  journal = {The Astrophysical Journal},
  volume = {923},
  number = {2},
  pages = {203},
  publisher = {The American Astronomical Society},
  issn = {0004-637X},
  doi = {10.3847/1538-4357/ac28f4},
  urldate = {2025-12-29},
  langid = {english},
}

@article{maWhyHighredshiftGalaxies2017,
  title = {Why Do High-Redshift Galaxies Show Diverse Gas-Phase Metallicity Gradients?},
  author = {Ma, Xiangcheng and Hopkins, Philip F. and Feldmann, Robert and Torrey, Paul and {Faucher-Gigu{\`e}re}, Claude-Andr{\'e} and Kere{\v s}, Du{\v s}an},
  year = 2017,
  month = may,
  journal = {Monthly Notices of the Royal Astronomical Society},
  volume = {466},
  number = {4},
  pages = {4780--4794},
  issn = {0035-8711},
  doi = {10.1093/mnras/stx034},
  urldate = {2025-12-29},
}

@article{rupkeGALAXYMERGERSMASS2010,
  title = {{{GALAXY MERGERS AND THE MASS}}--{{METALLICITY RELATION}}: {{EVIDENCE FOR NUCLEAR METAL DILUTION AND FLATTENED GRADIENTS FROM NUMERICAL SIMULATIONS}}},
  shorttitle = {{{GALAXY MERGERS AND THE MASS}}--{{METALLICITY RELATION}}},
  author = {Rupke, David S. N. and Kewley, Lisa J. and Barnes, Joshua E.},
  year = 2010,
  month = feb,
  journal = {The Astrophysical Journal Letters},
  volume = {710},
  number = {2},
  pages = {L156},
  publisher = {The American Astronomical Society},
  issn = {2041-8205},
  doi = {10.1088/2041-8205/710/2/L156},
  urldate = {2025-12-29},
  langid = {english},
}

@article{floresStarformingRegionsMetallicity2014,
  title = {Star-Forming Regions and the Metallicity Gradients in the Tidal Tails: The Case of {{NGC}}~92\ding{72}},
  shorttitle = {Star-Forming Regions and the Metallicity Gradients in the Tidal Tails},
  author = {{Torres-Flores}, S. and Scarano, S. and {Mendes de Oliveira}, C. and {de Mello}, D. F. and Amram, P. and Plana, H.},
  year = 2014,
  month = feb,
  journal = {Monthly Notices of the Royal Astronomical Society},
  volume = {438},
  number = {2},
  pages = {1894--1908},
  issn = {0035-8711},
  doi = {10.1093/mnras/stt2340},
  urldate = {2025-12-29}
}

@article{chiangGalaxyProtoclustersDrivers2017,
  title = {Galaxy {{Protoclusters}} as {{Drivers}} of {{Cosmic Star Formation History}} in the {{First}} 2 {{Gyr}}},
  author = {Chiang, Yi-Kuan and Overzier, Roderik and Gebhardt, Karl},
  year = 2017,
  month = aug,
  journal = {The Astrophysical Journal},
  volume = {844},
  number = {2},
  pages = {L23},
  issn = {0004-637X},
  doi = {10.3847/2041-8213/aa7e7b},
  urldate = {2024-03-15},
  langid = {english},
}

@article{shimakawaMAHALODeepCluster2018a,
  title = {{{MAHALO Deep Cluster Survey II}}. {{Characterizing}} Massive Forming Galaxies in the {{Spiderweb}} Protocluster at z = 2.2},
  author = {Shimakawa, Rhythm and Koyama, Yusei and R{\"o}ttgering, Huub J. A. and Kodama, Tadayuki and Hayashi, Masao and Hatch, Nina A. and Dannerbauer, Helmut and Tanaka, Ichi and Tadaki, Ken-ichi and Suzuki, Tomoko L. and Fukagawa, Nao and Cai, Zheng and Kurk, Jaron D.},
  year = 2018,
  month = dec,
  journal = {Monthly Notices of the Royal Astronomical Society},
  volume = {481},
  pages = {5630--5650},
  publisher = {OUP},
  issn = {0035-8711},
  doi = {10.1093/mnras/sty2618},
  urldate = {2024-06-19},
}

@article{shiSpectroscopicConfirmationTwo2021a,
  title = {Spectroscopic {{Confirmation}} of {{Two Extremely Massive Protoclusters}}, {{BOSS1244}} and {{BOSS1542}}, at z = 2.24},
  author = {Shi, Dong Dong and Cai, Zheng and Fan, Xiaohui and Zheng, Xian Zhong and Huang, Yun-Hsin and Xu, Jiachuan},
  year = 2021,
  month = jul,
  journal = {The Astrophysical Journal},
  volume = {915},
  number = {1},
  pages = {32},
  publisher = {The American Astronomical Society},
  issn = {0004-637X},
  doi = {10.3847/1538-4357/abfec0},
  urldate = {2025-12-31},
  langid = {english},
}

@article{caiDiscoveryEnormousLya2017,
  title = {Discovery of an {{Enormous Ly$\alpha$ Nebula}} in a {{Massive Galaxy Overdensity}} at z = 2.3},
  author = {Cai, Zheng and Fan, Xiaohui and Yang, Yujin and Bian, Fuyan and Prochaska, J. Xavier and Zabludoff, Ann and McGreer, Ian and Zheng, Zhen-Ya and Green, Richard and Cantalupo, Sebastiano and Frye, Brenda and Hamden, Erika and Jiang, Linhua and Kashikawa, Nobunari and Wang, Ran},
  year = 2017,
  month = mar,
  journal = {The Astrophysical Journal},
  volume = {837},
  number = {1},
  pages = {71},
  publisher = {The American Astronomical Society},
  issn = {0004-637X},
  doi = {10.3847/1538-4357/aa5d14},
  urldate = {2025-12-31},
  langid = {english},
}

@article{caiMAPPINGMOSTMASSIVE2016,
  title = {{{MAPPING THE MOST MASSIVE OVERDENSITY THROUGH HYDROGEN}} ({{MAMMOTH}}). {{I}}. {{METHODOLOGY}}},
  author = {Cai, Zheng and Fan, Xiaohui and Peirani, Sebastien and Bian, Fuyan and Frye, Brenda and McGreer, Ian and Prochaska, J. Xavier and Lau, Marie Wingyee and Tejos, Nicolas and Ho, Shirley and Schneider, Donald P.},
  year = 2016,
  month = dec,
  journal = {The Astrophysical Journal},
  volume = {833},
  number = {2},
  pages = {135},
  publisher = {The American Astronomical Society},
  issn = {0004-637X},
  doi = {10.3847/1538-4357/833/2/135},
  urldate = {2025-12-31},
  langid = {english},
}

@article{zhengMAMMOTHConfirmationTwo2021,
  title = {{{MAMMOTH}}: Confirmation of Two Massive Galaxy Overdensities at z = 2.24 with {{H$\alpha$}} Emitters},
  shorttitle = {{{MAMMOTH}}},
  author = {Zheng, Xian Zhong and Cai, Zheng and An, Fang Xia and Fan, Xiaohui and Shi, Dong Dong},
  year = 2021,
  month = jan,
  journal = {Monthly Notices of the Royal Astronomical Society},
  volume = {500},
  number = {4},
  pages = {4354--4364},
  issn = {0035-8711},
  doi = {10.1093/mnras/staa2882},
  urldate = {2025-12-31},
  langid = {english},
}

@ARTICLE{yangMAMMOTHGrismRevisitingMassMetallicity2025,
       author = {{Yang}, Yiming and {Wang}, Xin and {He}, Xianlong and {Tsai}, Chao-Wei and {Cai}, Zheng and {Li}, Zihao and {Malkan}, Matthew A. and {Shi}, Dong Dong and {Alavi}, Anahita and {Bian}, Fuyan and {Colbert}, James and {Fan}, Xiaohui and {Henry}, Alaina L. and {Teplitz}, Harry I. and {Zheng}, Xian Zhong},
        title = "{MAMMOTH-Grism: Revisiting the Mass─Metallicity Relation in Protocluster Environments at Cosmic Noon}",
      journal = {\apj},
         year = 2026,
        month = jan,
       volume = {997},
       number = {1},
          eid = {95},
        pages = {95},
          doi = {10.3847/1538-4357/ae27cf},
archivePrefix = {arXiv},
       eprint = {2511.16154},
 primaryClass = {astro-ph.GA},
       adsurl = {https://ui.adsabs.harvard.edu/abs/2026ApJ...997...95Y}
}

@article{sandersMOSDEFSurveyEvolution2021,
  title = {The {{MOSDEF Survey}}: {{The Evolution}} of the {{Mass}}--{{Metallicity Relation}} from z = 0 to z {$\sim$} 3.3*},
  shorttitle = {The {{MOSDEF Survey}}},
  author = {Sanders, Ryan L. and Shapley, Alice E. and Jones, Tucker and Reddy, Naveen A. and Kriek, Mariska and Siana, Brian and Coil, Alison L. and Mobasher, Bahram and Shivaei, Irene and Dav{\'e}, Romeel and Azadi, Mojegan and Price, Sedona H. and Leung, Gene and Freeman, William R. and Fetherolf, Tara and de Groot, Laura and Zick, Tom and Barro, Guillermo},
  year = 2021,
  month = jun,
  journal = {The Astrophysical Journal},
  volume = {914},
  number = {1},
  pages = {19},
  publisher = {The American Astronomical Society},
  issn = {0004-637X},
  doi = {10.3847/1538-4357/abf4c1},
  urldate = {2024-09-27},
  langid = {english},
}

@article{yuanSystematicsMetallicityGradient2013,
       author = {{Yuan}, T.-T. and {Kewley}, L.~J. and {Rich}, J.},
        title = "{Systematics in Metallicity Gradient Measurements. I. Angular Resolution, Signal to Noise, and Annular Binning}",
      journal = {\apj},
         year = 2013,
        month = apr,
       volume = {767},
       number = {2},
          eid = {106},
        pages = {106},
          doi = {10.1088/0004-637X/767/2/106},
archivePrefix = {arXiv},
       eprint = {1302.6232},
 primaryClass = {astro-ph.CO},
       adsurl = {https://ui.adsabs.harvard.edu/abs/2013ApJ...767..106Y}
}

@article{daikuharaAssociationColdGas2025a,
  title = {Association of Cold Gas, Massive Galaxies, and {{AGNs}} in a Filamentary Protocluster Traced by Triple Narrow-Band Imaging},
  author = {Daikuhara, Kazuki and Kodama, Tadayuki and Kusakabe, Haruka and Steidel, Charles C and Tanaka, Ichi and Kikuta, Satoshi and Umehata, Hideki and Shimakawa, Rhythm and Koyama, Yusei and Motohara, Kentaro and Konishi, Masahiro and {Perez-Martinez}, Jose Manuel and Kubo, Mariko and Erb, Dawn and Takahashi, Kosuke and Fukushima, Keita},
  year = 2025,
  month = dec,
  journal = {Monthly Notices of the Royal Astronomical Society},
  volume = {544},
  number = {2},
  pages = {2365--2386},
  issn = {0035-8711},
  doi = {10.1093/mnras/staf1772},
  urldate = {2026-01-02}
}

@article{wangEnvironmentalDependenceMass2023,
  title = {Environmental {{Dependence}} of the {{Mass}}--{{Metallicity Relation}} in {{Cosmological Hydrodynamical Simulations}}},
  author = {Wang, Kai and Wang, Xin and Chen, Yangyao},
  year = 2023,
  month = jul,
  journal = {The Astrophysical Journal},
  volume = {951},
  number = {1},
  pages = {66},
  publisher = {The American Astronomical Society},
  issn = {0004-637X},
  doi = {10.3847/1538-4357/acd633},
  urldate = {2026-01-02},
  langid = {english},
}

@article{jonesGRISMLENSAMPLIFIEDSURVEY2015,
  title = {{{THE GRISM LENS-AMPLIFIED SURVEY FROM SPACE}} ({{GLASS}}). {{II}}. {{GAS-PHASE METALLICITY AND RADIAL GRADIENTS IN AN INTERACTING SYSTEM AT Z}} {$\simeq$} 2},
  author = {Jones, T. and Wang, X. and Schmidt, K. B. and Treu, T. and Brammer, G. B. and Brada{\v c}, M. and Dressler, A. and Henry, A. L. and Malkan, M. A. and Pentericci, L. and Trenti, M.},
  year = 2015,
  month = feb,
  journal = {The Astronomical Journal},
  volume = {149},
  number = {3},
  pages = {107},
  publisher = {The American Astronomical Society},
  issn = {1538-3881},
  doi = {10.1088/0004-6256/149/3/107},
  urldate = {2026-02-05},
  langid = {english},
}

@article{jonesMEASUREMENTMETALLICITYGRADIENT2010,
  title = {{{MEASUREMENT OF A METALLICITY GRADIENT IN A}} z = 2 {{GALAXY}}: {{IMPLICATIONS FOR INSIDE-OUT ASSEMBLY HISTORIES}}},
  shorttitle = {{{MEASUREMENT OF A METALLICITY GRADIENT IN A}} z = 2 {{GALAXY}}},
  author = {Jones, Tucker and Ellis, Richard and Jullo, Eric and Richard, Johan},
  year = 2010,
  month = dec,
  journal = {The Astrophysical Journal Letters},
  volume = {725},
  number = {2},
  pages = {L176},
  publisher = {The American Astronomical Society},
  issn = {2041-8205},
  doi = {10.1088/2041-8205/725/2/L176},
  urldate = {2026-02-05},
  langid = {english},
}

@article{oppenheimerMassMetalEnergy2008,
  title = {Mass, Metal, and Energy Feedback in Cosmological Simulations},
  author = {Oppenheimer, Benjamin D. and Dav{\'e}, Romeel},
  year = 2008,
  month = jun,
  journal = {Monthly Notices of the Royal Astronomical Society},
  volume = {387},
  number = {2},
  pages = {577--600},
  issn = {0035-8711, 1365-2966},
  doi = {10.1111/j.1365-2966.2008.13280.x},
  urldate = {2024-10-15},
  langid = {english},
}

@article{el-badryGasKinematicsMorphology2018,
  title = {Gas Kinematics, Morphology and Angular Momentum in the {{FIRE}} Simulations},
  author = {{El-Badry}, Kareem and Quataert, Eliot and Wetzel, Andrew and Hopkins, Philip F. and Weisz, Daniel R. and Chan, T. K. and Fitts, Alex and {Boylan-Kolchin}, Michael and Kere{\v s}, Du{\v s}an and {Faucher-Gigu{\`e}re}, Claude-Andr{\'e} and {Garrison-Kimmel}, Shea},
  year = 2018,
  month = jan,
  journal = {Monthly Notices of the Royal Astronomical Society},
  volume = {473},
  number = {2},
  pages = {1930--1955},
  issn = {0035-8711},
  doi = {10.1093/mnras/stx2482},
  urldate = {2024-10-15},
}

@article{zhouMAMMOTHMOSFIREEnvironmentalEffects2025a,
  title = {{{MAMMOTH-MOSFIRE}}: {{Environmental Effects}} on {{Galaxy Interstellar Medium}} at z {$\sim$} 2},
  shorttitle = {{{MAMMOTH-MOSFIRE}}},
  author = {Zhou, Hang and Wang, Xin and Malkan, Matthew A. and Treu, Tommaso and Yang, Yiming and Cai, Zheng and Fan, Xiaohui and Ju, Mengting and Shi, Dong Dong and Alavi, Anahita and Bian, Fuyan and Colbert, James and Henry, Alaina L. and Li, Sijia and Li, Zihao and Teplitz, Harry I. and Zhan, Hu and Zheng, Xian Zhong and Zheng, Zheng and Jin, Yifei},
  year = 2025,
  month = nov,
  journal = {The Astrophysical Journal},
  volume = {993},
  number = {2},
  pages = {231},
  publisher = {The American Astronomical Society},
  issn = {0004-637X},
  doi = {10.3847/1538-4357/ae0649},
  urldate = {2026-02-05},
  langid = {english},
}

@ARTICLE{2025arXiv251008997S,
       author = {{Sun}, Xunda and {Wang}, Xin and {Jiang}, Fangzhou and {Mo}, Houjun and {Ho}, Luis C. and {Zhou}, Qianqiao and {Ma}, Xiangcheng and {Zhan}, Hu and {Wetzel}, Andrew and {Graf}, Russell L. and {Hopkins}, Philip F. and {Keres}, Dusan and {Stern}, Jonathan},
        title = "{Galaxy Metallicity Gradients in the Reionization Epoch from the FIRE-2 Simulations}",
      journal = {arXiv e-prints},
         year = 2025,
        month = oct,
          eid = {arXiv:2510.08997},
        pages = {arXiv:2510.08997},
          doi = {10.48550/arXiv.2510.08997},
archivePrefix = {arXiv},
       eprint = {2510.08997},
 primaryClass = {astro-ph.GA},
       adsurl = {https://ui.adsabs.harvard.edu/abs/2025arXiv251008997S},
}

@software{2019ascl.soft05001B,
       author = {{Brammer}, Gabe},
        title = "{Grizli: Grism redshift and line analysis software}",
 howpublished = {Astrophysics Source Code Library, record ascl:1905.001},
         year = 2019,
        month = may,
          eid = {ascl:1905.001},
archivePrefix = {ascl},
       eprint = {1905.001},
       adsurl = {https://ui.adsabs.harvard.edu/abs/2019ascl.soft05001B}
}
\bibliographystyle{aasjournal}



\end{CJK*}
\end{document}